\documentclass[10pt]{iopart}

\usepackage[utf8]{inputenc}
\usepackage{graphicx}
\usepackage[breaklinks=true,colorlinks=true,linkcolor=blue,urlcolor=
blue,citecolor=blue]{hyperref}
\usepackage{bigfoot}
\usepackage{pdflscape}
\usepackage{booktabs}
\usepackage{array}
\graphicspath{{pics/}}
\usepackage{longtable}
\usepackage{url}
\usepackage{soul}

\begin{document}

\title[Barrier discharges in CO$_2$ - $E/N$ determination]{Barrier discharges in CO$_2$ - optical emission spectra analysis and {\it E/N} determination from intensity ratio}

\author{Tom\'a\v s Hoder$^{1,\ast}$, David Prokop$^{1}$, Corentin Bajon$^{2}$,\\ Simon Dap$^{2}$, Zden\v ek Navr\'atil$^{1}$ and Nicolas Naud\'e$^{2}$}

\ead{$^{\ast}$hoder@physics.muni.cz}

\address{$^{1}$Department of Plasma Physics and Technology, Faculty of Science, Masaryk University,
	Kotl\'a\v{r}sk\'a 2, 611 37 Brno, Czech Republic}
	
	\address{$^{2}$LAPLACE, Universit\'e de Toulouse, CNRS, INPT, UPS, Toulouse, France}

\date{\today}

\begin{abstract}
We study barrier discharges in CO$_2$ using optical emission spectroscopy and electrical measurements. We record the spectra of weak CO$_2$ barrier discharge emission in UV-VIS-NIR range. 
The CO spectral bands of 3$^{\mathrm{rd}}$ positive 
, {\AA}ngström 
and Asundi 
systems are identified in the spectra, together with the CO$_2^+$ bands of the UV-doublet 
and the Fox-Duffendack-Baker 
systems. 
The effect of CO$_2$ flow rate variations on the relative intensities of the CO and CO$_2$ optical emissions is also investigated. 
We analyze the optical emission of selected vibrational spectral bands and attempt to determine the $E/N$ using intensity ratio method with kinetic data available in the literature. 
The method, using intensities from different spectral bands, is tested on sinusoidal driven atmospheric pressure Townsend (APTD) and single-filament barrier discharges at atmospheric pressure. 
For APTD at different flows and known CO conversion factors, an attempt is made to use the ratio of selected CO to CO$_2$ band intensities. 
The technique of time-correlated single photon counting is used to obtain necessary sub-nanosecond temporal resolution and, importantly,  high signal sensitivity for the single-filament experiment and its utilization gives a promising result, $E/N$ of 330\,Td in barrier discharge streamer. 
Nevertheless, the uncertainties coming from different kinetic data 
and the low sensitivity of some rate coefficient ratios complicate an easy utilisation of the method. 
Future steps are therefore proposed, including uncertainty quantification and sensitivity analysis of the simplified model and necessary CO$_2^+$ and CO synthetic spectra fitting.

\end{abstract}

\maketitle

\newpage

\ioptwocol

\section{\label{intro} Introduction}

Dielectric barrier discharge is a widely applied scalable source of non-thermal low temperature plasma \cite{kogelschatz2003,brandenburg2017}. Its application spans from ozone generation, gas purification, surface treatment to UV-radiation generation or plasma-chemical synthesis and flow control \cite{eliasson1987,kozlov2000,kraus2002,corke2010,cernak2011,brandt2016}. 
Recently, the interest in CO$_2$ plasmas increased also due to environmental purposes \cite{pietanza2021} and the CO$_2$ plasma was investigated in numerous discharges, including the dielectric barrier ones \cite{brehmer2014,belov2016,ponduri2016,brandenburg2017b,douat2023,bajon2023}. 
The use of such plasma for CO$_2$ conversion is intensively studied as well \cite{brehmer2014,snoeckx2017}.
Importantly, the CO$_2$ may replace the significantly more potent greenhouse gas SF$_6$ in high-voltage circuit breakers where the understanding of ionization processes (including and especially pre-breakdown streamers) is of high interest \cite{seeger2015,seeger2017,rabie2018,vass2021}. It may become an important component of the mixture including the new SF$_6$ alternatives \cite{rankovic2020,vemu2023}. 
In all these cases, the detailed knowledge on pre-breakdown or partial-breakdown phenomena is needed and a suitable diagnostics may help to clarify the underlying processes or validate more complex numerical models.

The local electric field or mean electron energy are important plasma parameters for detailed description of the investigated plasmas and optical emission spectroscopy may be a suitable (non-invasive and in-situ) tool for their determination \cite{goldberg2022}. 
Typically, in nitrogen containing plasmas the intensity ratio method using spectral bands of molecular nitrogen is widely used in various plasmas \cite{kozlov2001,gharib2017,bilek2019,jansky2021,dijcks2022}, combinations of argon and nitrogen \cite{goekce2016} were proposed as well as developments of such method in pure argon under different conditions \cite{siepa2014,dyatko2021,kusyn2023}. The method is based on a simple collision-radiative model (CRM). 
The intensity of emitted light from pre-breakdown or partial discharges in CO$_2$ is known to be rather weak, yet for special cases - e.g. a  reproducible single-filament barrier discharges - accumulative methods may enable a study of specific spectral signatures in close detail. In such case, the localised and repetitive barrier discharge plasma serves as a simplified experimental model for more complicated application-oriented arrangements and it is well suitable for a validation of numerical models, similarly as it was done for example for barrier discharge in air \cite{jansky2021}. 

The time-correlated single photon counting (TCSPC) technique for enhanced optical emission spectroscopy is certainly one of the accumulative methods suitable in such case. Since its first use in the field of low temperature plasma physics, it has offered a high temporal resolution and a huge dynamic range, enabling to study ultra-fast plasma physical processes of high optical emission intensity and slow weakly emitting discharge events at the same time. Study of the streamer and the Townsend phases of dielectric barrier discharge development is an excellent example of its applicability. 
Earlier, the technique was used for rf-plasma investigation \cite{rosny1983,kokubo1990}, to study the Trichel pulses in negative corona discharge \cite{ushita1967,gravendeel1988}, positive streamers \cite{kondo1980,shcherbakov2007} and barrier discharges \cite{kozlov1995,kozlov2001}. Recently, the investigations using TCSPC resulted in spatiotemporal electric field quantification in subsequent streamers in volume barrier discharge \cite{jahanbakhsh2019correlation}, transient spark discharges \cite{janda2017} or in surface streamers in coplanar barrier discharge, in good agreement with results of computer simulations \cite{jansky2021}.

In this article, we study optical emission spectra of homogeneous and filamentary barrier discharges in CO$_2$. 
Investigating the homogeneous Townsend barrier discharge (atmospheric pressure Townsend discharge, APTD, see \cite{bajon2023}) and filamentary barrier discharges in CO$_2$, we identify the optical emission spectra based on the known literature, attempt to clarify the differences and propose diagnostic methods for $E/N$ determination based on intensity ratios of molecular spectral bands of CO or CO$_2^+$. 
The APTD is investigated using intensified CCD (ICCD) cameras and using the known CO$_2$ to CO conversion factors $\alpha$ we also combine the intensity ratios of CO$_2^+$ and CO bands for diagnostic purposes.
The optical emission of single-filament dielectric barrier discharge in CO$_2$ is studied at sub-nanosecond timescales using the TCSPC and we use the discharge as an experimental model for development of a spectroscopic method for plasma diagnostics. 
Based on the obtained spectroscopic data and their evaluation using simplified kinetic model with different electron excitation cross-sections, excitation Franck-Condon factors, radiative lifetimes or collision quenching rates, we discuss the applicability of developed methods and propose further steps to be done for improvement of the methods' accuracy. 
We also compare the obtained values of experimentally accessed $E/N$ to those known from literature, e.g. \cite{ponduri2016}. 
Using electrical current measurements we quantify the transferred electrical charge within a single micro-discharge and compare it also to the APTD in CO$_2$ and known experimental results \cite{belov2016,ponduri2016,brandenburg2017b,douat2023}. 

The article is structured in the following way. In the next section, the experimental setup and diagnostic methods are described together with the overall electrical characterization of investigated CO$_2$ discharges. In the third section, the optical emission spectra of investigated barrier discharges are analysed and discussed. The general approach for development of the intensity ratio method is described in detail and based on the previous discussion a simplified kinetic schemes for $E/N$ determination both from CO and CO$_2^+$ spectral bands' intensities are proposed in the fourth section. In the fifth section, the results of the TCSPC measurement as well as APTD ICCD spectra are presented for selected spectral signatures, analyzed using the proposed models and discussed in closer detail. We summarize the findings in the final section.

\begin{figure}[hbt]
\begin{center}
a)
\includegraphics[clip = true,width=0.3\columnwidth]{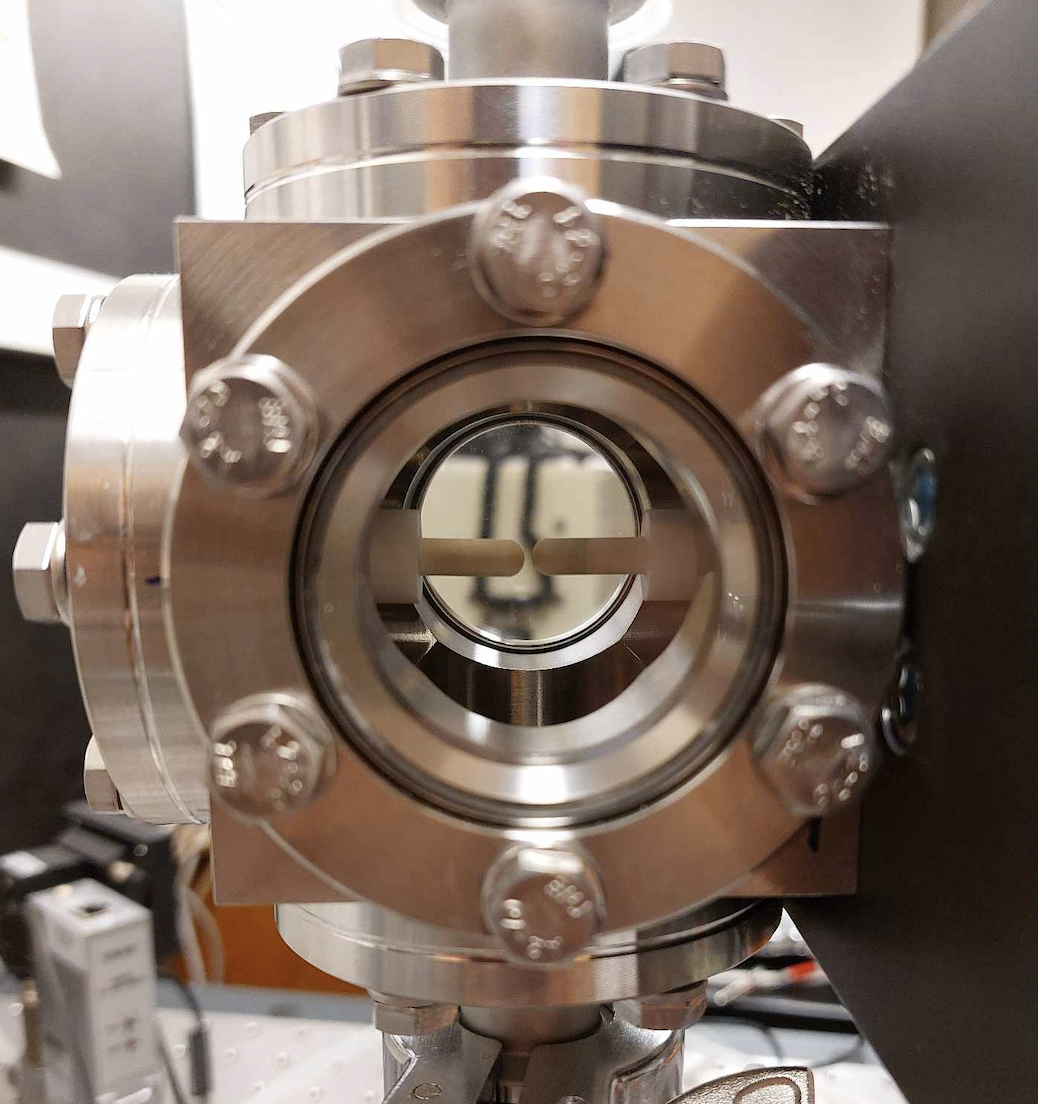}
b)
\includegraphics[clip = true,width=0.126\columnwidth]{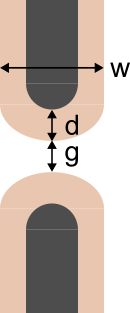}
c)
\includegraphics[clip = true,width=0.15\columnwidth]{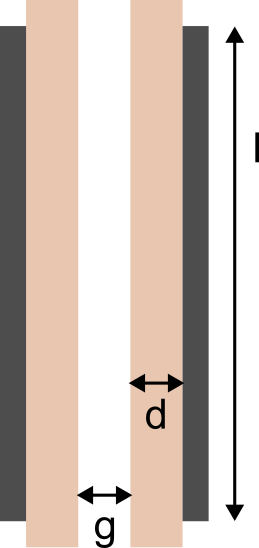}
\caption{Stainless steel vacuum chamber with two dielectrics-covered electrodes inside used for single-filament barrier discharge experiments, a). The schematic description of electrode arrangements for the single-filament discharge is shown in part b), where $d$ = $g$ = 1\,mm and $w$ = 5\,mm. Dielectrics is shown using beige color and conductive metal using dark grey. In the part c), the arrangement for the APTD and the multi-filament discharge is shown, with $g$ = 1\,mm, length of the square electrode of $l$ = 3\,cm and $d$ = 1\,mm.}
\label{chamber}
\end{center}
\end{figure}

\section{\label{experiment} Experimental setup and electrical parameters of barrier discharges}

In this article, we analyze electrical signatures and optical emission spectra (OES) from discharges generated using two electrode arrangements (see Fig.\,\ref{chamber}) and operated in two different regimes: filamentary and homogeneous. In total, barrier discharges in (i) single-filament (using setup in Fig.\,\ref{chamber}b), (ii) multi-filament and (iii) APTD mode (both using setup Fig.\,\ref{chamber}c) were investigated.

The homogeneous and multi-filament discharges were driven by a low-frequency generator Keysight 33210A connected to an audio amplifier and a high-voltage transformer. 
Electrical signals were measured using a high-voltage probe at the output of the transformer (Tektronix P6015A) and a probe (Rohde \& Schwarz RT-ZP03) over a 1\,k$\Omega$ shunt resistor in series with the discharge cell.
Both signals were recorded using oscilloscope Rohde \& Schwarz RTB2004, bandwidth 300\,MHz. 
CO$_2$ Alphagaz\,1 from Air Liquide with a purity of 99.9\% was used.
The optical emission spectra in barrier discharges were measured using a spectrometer Acton SP2500 from Princeton Instruments with monochromator 
coupled with a PI-MAX 3 camera (Gen III Filmless Unigen II intensifier) for wavelength above 440\,nm. 
The slit of the spectrometer was set at 200\,$\mu$m aperture. 
The spectra were calibrated using an IntelliCal Wavelength Calibration Source from Princeton Instruments and a UV-visible Halogen Deuterium lamp DH-3-BAL-CAL from Ocean Optics for intensity calibration. More details can be found in \cite{bajon2023}. 
For spectra under 440\,nm the ICCD camera PI-MAX 2 (Gen II RB Fast Gate intensifier) was used as it is very sensitive in interval from 250 to 500\,nm, together with monochromator Acton SP2500 with  150\#/mm grating. The slit aperture was set to 600\,$\mu$m. These spectral intervals are marked in the respective figures.

\begin{figure}[hbt]
\begin{center}
\includegraphics[clip = true,width=0.95\columnwidth]{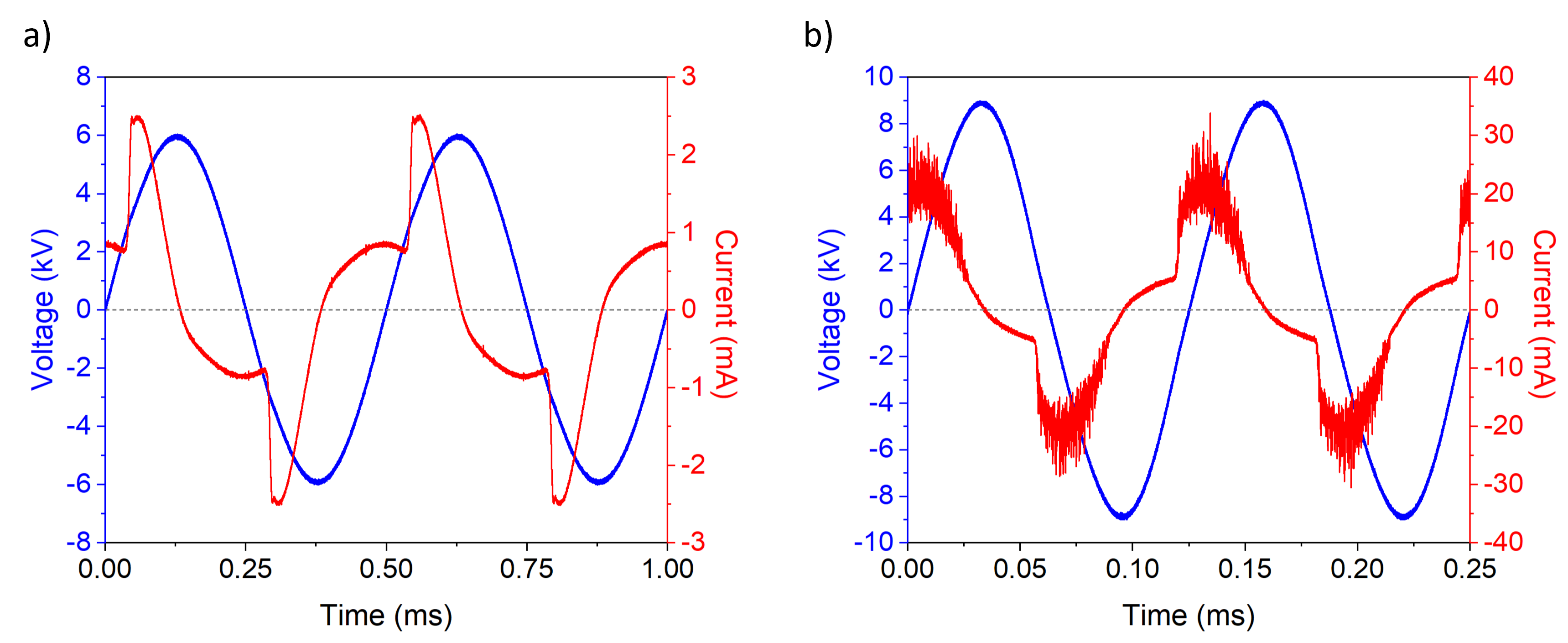}
\caption{Oscillograms of a) an APTD (2\,kHz 12\,kV$_{pp}$) and a multi-filamentary barrier discharge (8\,kHz 18\,kV$_{pp}$) in CO$_2$.}
\label{electro1}
\end{center}
\end{figure}

The homogeneous APTD regime was introduced in our previous article in \cite{bajon2023}, where also the electrode arrangement was described: consisting of two 115$\times$70\,mm parallel alumina plates with a square electrode of 3$\times$3\,cm$^2$ painted on both plates using a silver-based conductor ESL 9916, the thickness of each alumina plate was 1\,mm and the gap was 1\,mm, too. Here, the APTD was generated using applied voltage of sinusoidal waveform with frequency of 2\,kHz and with peak-to-peak value of 12\,kV$_{pp}$. 
The voltage and current waveforms are shown in Fig.\,\ref{electro1}a). Only one broad current peak (hump of few hundreds microseconds) occurs during each half-period with amplitude of approximately 1.5\,mA, if the displacement current is subtracted. 
In the same larger-scale arrangement the multi-filamentary barrier discharge was operated as well, under following conditions: using applied voltage of sinusoidal waveform with frequency of 8\,kHz and with peak-to-peak value of 18\,kV$_{pp}$. 
The voltage and current waveforms are shown in Fig.\,\ref{electro1}b). Multiple current peaks occurs with amplitude up to approximately 20\,mA, if the displacement current is subtracted. 

\begin{figure}[hbt]
\begin{center}
\includegraphics[clip = true,width=0.45\columnwidth]{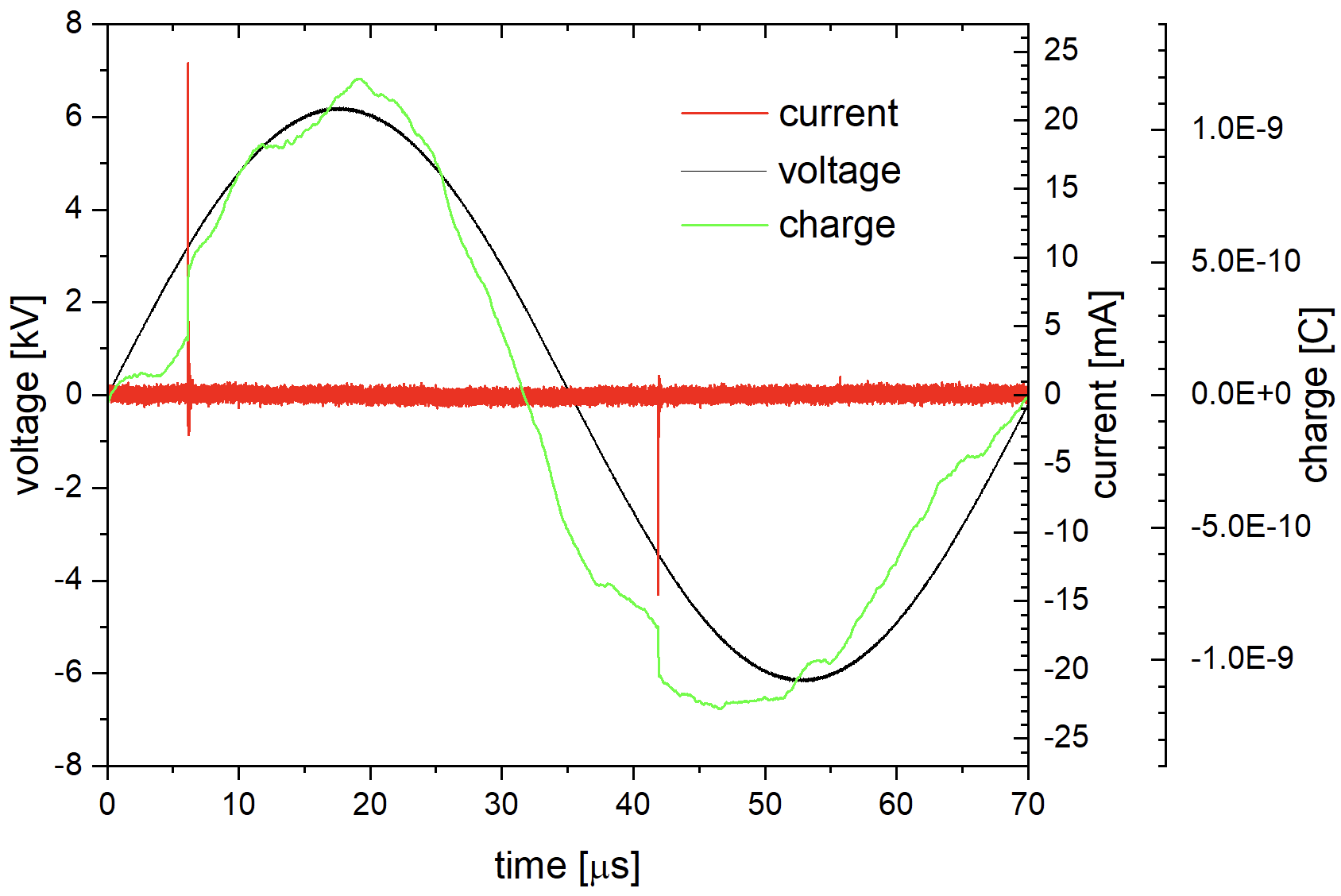}
\caption{Electrical measurements of the single-filament discharge in atmospheric pressure CO$_2$. The single current peaks have similar amplitudes like in \cite{brandenburg2017b} and \cite{bajon2023} where filamentary barrier discharges were investigated under similar conditions as well.}
\label{electro2}
\end{center}
\end{figure}

The single-filament mode of the barrier discharge was driven by a sinusoidal applied voltage waveform of frequency 14\,kHz and with a peak-to-peak value of 12.5\,kV$_{pp}$. The vacuum chamber and the electrode arrangement are shown in Fig.\,\ref{chamber}a) and b), respectively. 
Before every experiment, the vacuum chamber was evacuated down to 2\,mbar and then filled with CO$_2$ of 99.99\% purity and flushed with flow rate of 0.5\,slm during the whole experiment. 
All experiments presented here were done at atmospheric pressure. The voltage was measured using the Tektronix P6015A probe and the electrical current using the so called BNC probe (voltage drop measurements on, in total, 50\,Ohm resistor connected as star-network in coaxial cable), as described in \cite{synek2018}. 
Electrical waveforms were recorded on a high-definition oscilloscope, 
Keysight DSO-2 204A (band-width 6\,GHz, 10-bit ADC, 20\,GSa). 
The typical current and voltage waveforms, together with the transferred charge, are shown in Fig.\,\ref{electro2}. 
Similarly as in already presented results on filamentary barrier discharges in CO$_2$ under comparable conditions, the filaments have amplitude of approx. 20 to 30\,mA, duration around 10\,ns and transferred charge of around 0.3\,nC, see \cite{brandenburg2017b,bajon2023}.  Here, the emission spectra of single-filament discharge were also measured using ICCD camera Pimax\,4 (PM4-1024i-RB-PS-18-P46, Princeton Instruments) and a monochromator ARC SpectraPro-2500i with gratings 1200\#/mm and 2400\#/mm, mostly to assist the temporally resolved spectra measurements.

The temporally high-resolution spectra of the single-filament discharge were measured using the TCSPC technique. 
This technique substitutes the real-time measurement of the discharge light emission waveform by a statistically averaged determination of the cross-correlation function between two optical signals, both originating from the same source - the reproducible discharge itself \cite{kozlov2001}. 
These two signals are the so-called main (spatially and spectrally resolved signal with single photon resolution) and the synchronization signal. 
The time between the synchronization and the main signal is measured. 
Both signals were focused to the monochromator (main) or directly to the PMT (synchronization) using a two-inch lenses. 
TCSPC was utilized using a SPC-150 module from Becker\&Hickl GmbH. Two photomultipliers (PMTs) PMC-100-20 and PMC-100-4 (both Becker\&Hickl GmbH) were used as a detectors for the main and the synchronization signals, as it was done in \cite{kusyn2023}. 
The TCSPC is an accumulation technique and usually 10$^4$ correlations are needed to reconstruct the original light pulse properly. The transit time spread of the used PMTs for these measurements was 180\,ps. 

\begin{figure}[hbt]
\begin{center}
\includegraphics[clip = true,width=0.5\columnwidth]{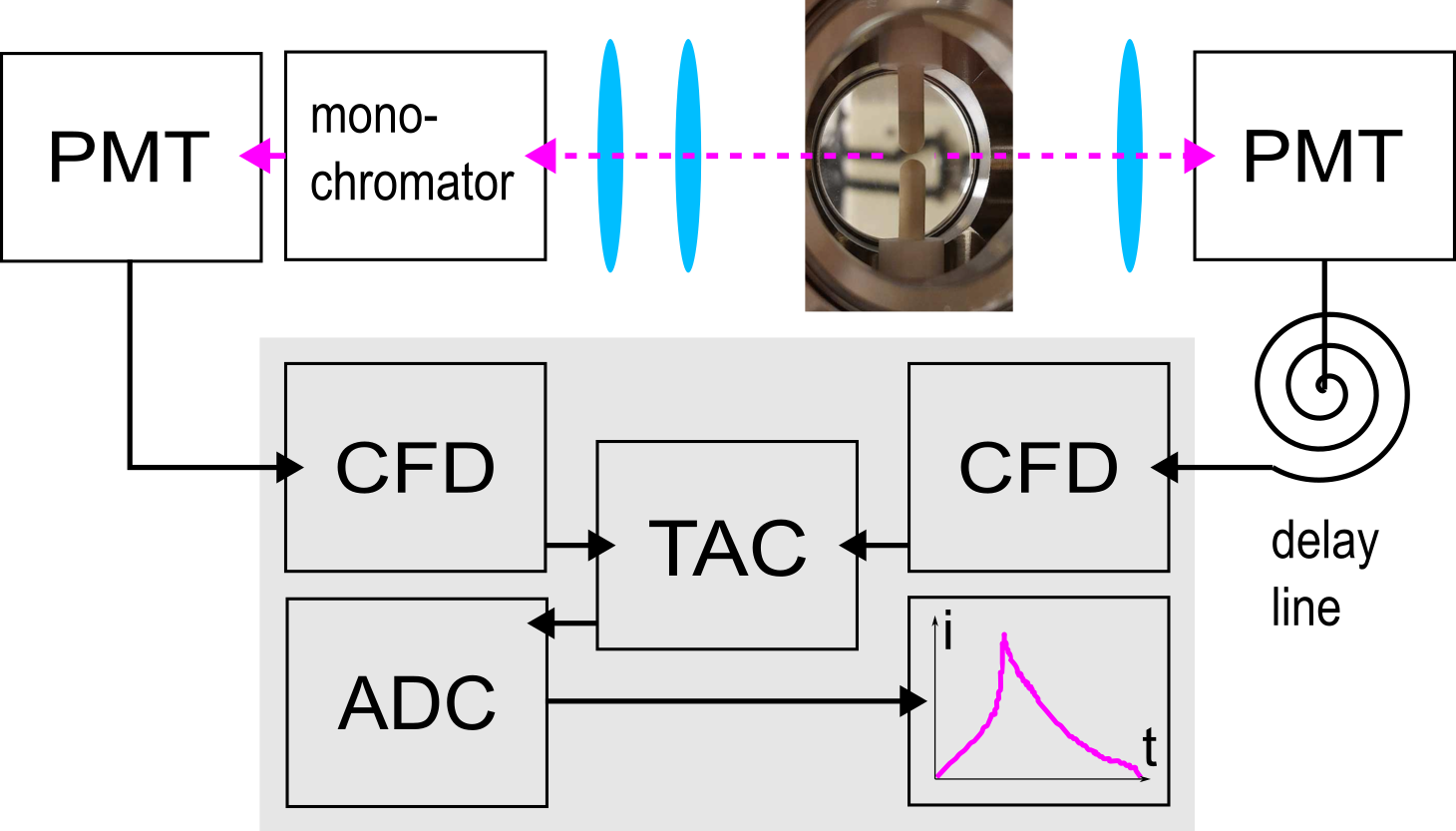}
\caption{Experimental setup used for the TCSPC measurements. The main signal path is going through the monochromator, the synchronization to the right direction from the discharge. The grey block designates the internal parts of the single photon counting module.}
\label{setup}
\end{center}
\end{figure}

The scheme of the TCSPC setup is shown in Fig.\,\ref{setup}. The PMT denotes photomultiplier, CFD is constant fraction discriminator, TAC time to amplitude converter and ADC analogue to digital converter. CFD, TAC and ADC high-frequency electronics creates the TCSPC module responsible for the time-correlated photon counting. 
The main signal path is visible to the left of the discharge and the synchronization (or reference) signal to the right. 
The same setup was used as in \cite{kusyn2023}, with slightly differing internal parameters of the TCSPC module due to the significantly lower signal intensity in this case if compared to the argon discharge as investigated in \cite{kusyn2023}. 
For more details on the TCSPC technique see \cite{kozlov2001,becker2005}. 

When using the TCSPC technique for intensity measurements, only a certain spectral interval is taken into account. For given linear dispersion of the setup and slit opening it gives a spectral resolution, in our case about 0.46\,nm. The spectral resolution was measured independently using a calibration lamp. The overview spectra around the spectral band head is recorded using the TCSPC as well. The intensity is then computed as an integral, from the measured fraction of the whole band intensity.

\section{\label{secX} Optical emission spectra in barrier discharges in CO$_2$}

The optical emission from the low-temperature plasma of dielectric barrier discharge (DBD) in close-atmospheric pressure CO$_2$ gas is typically very weak. 
The spectrum consists of multiple molecular spectral systems and atomic lines. 
The composition and relative intensities of the spectral signatures are dependent on the physical mechanism responsible for the neutral gas ionization and excitation - different for APTD or filamentary discharge. It can also change due to different applied power or gas flow rate. We will discuss this later.

\begin{figure*}[hbt]
\begin{center}
\includegraphics[clip = true,width=1.8\columnwidth]{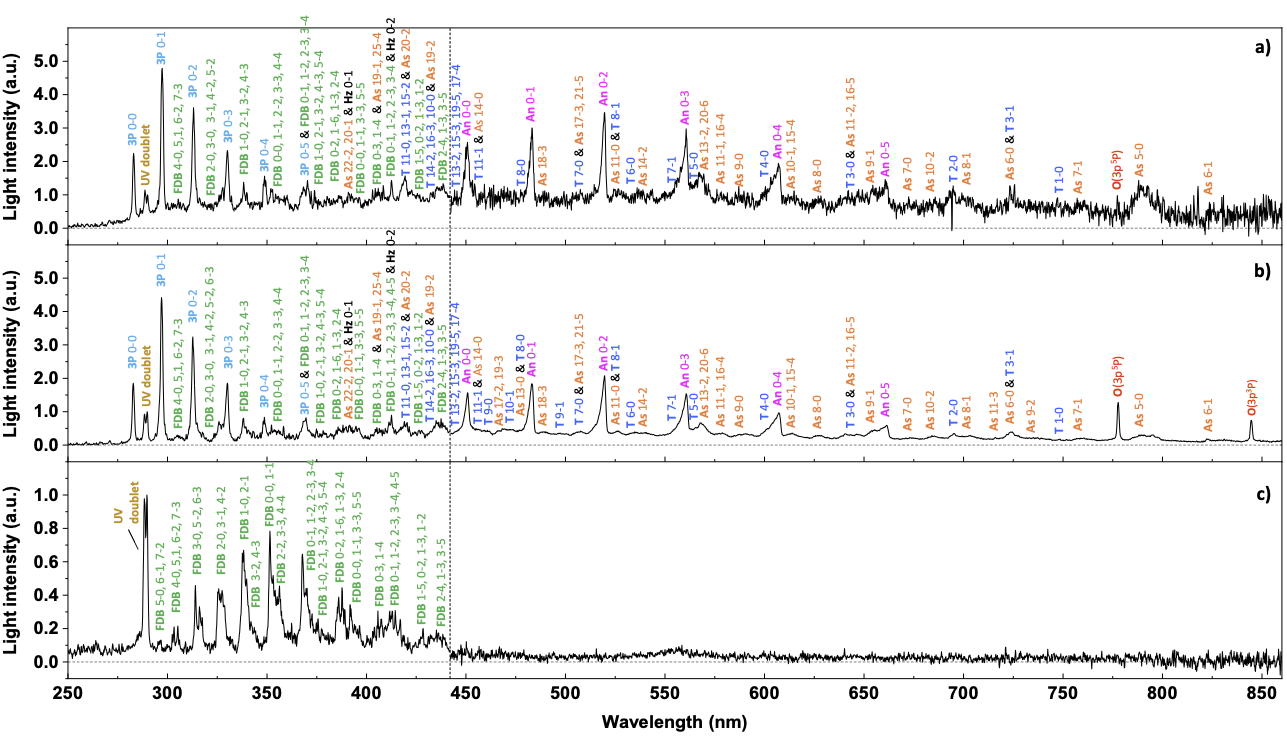}
\caption{Comparison of the OES spectra from a) APTD (1\,mm gap, 2\,kHz, 12\,kV$_{pp}$), b) multi-filament BD (1\,mm gap, 8\,kHz, 18\,kV$_{pp}$) and c) single-filament BD (1\,mm gap, 14.23\,kHz, 13.3\,kV$_{pp}$) in atmospheric pressure CO$_2$. The denoted spectra are systems of UV-doublet (brown), Fox-Duffendack-Baker (FDB, green) both of CO$_2^+$, 3$^{\mathrm{rd}}$ positive system of CO (3P, light blue), {\AA}ngstr{\"o}m (An, magenta), Asundi (As, orange), Herzberg (Hz, black), Triplet (T, violet) and oxygen atomic lines (red). In the case of low signal, the spectral descriptions serve for possible orientation rather than exact identification. }
\label{comparison}
\end{center}
\end{figure*}

It was shown by Bajon et al.\,\cite{bajon2023} that the spectra of barrier discharges in atmospheric pressure CO$_2$ in larger-scale multi-filamentary or in homogeneous APTD mode can significantly differ. 
Here, we analyze in closer detail these discharges and their spectra, and add the sinusoidal driven barrier discharge operated in single-filament mode for further comparison. 
All mentioned spectra are shown in Fig.\,\ref{comparison} for comparison. As it was done in \cite{bajon2023} we used a spectrometer Acton SP2500 with 300\#/mm grating. 
The spectra were calibrated by dividing the intensity by the acquisition time which is equal to the number of accumulations multiplied by the exposure time. 
The left and right part of the spectrum for given discharge type is normalized by dividing the intensity by the intensity of the spectral band with vibronic transition (0$\rightarrow$1) of the {\AA}ngstr{\"o}m system, to allow both part to be comparable. 
All three spectra were normalized onto the UV doublet intensity. 
Note, that the frequency for the single-filament discharge for this spectra is slightly different than for the TCSPC measurements, nevertheless no differences in OES were identified.

The spectra of the APTD and the multi-filamentary discharge are similar. 
One of the most intensive is the emission of the {\AA}ngstr{\"o}m system of the CO molecule:

\begin{equation}
\label{angstrom}
\mathrm{CO(B}^1\Sigma^+) \rightarrow \mathrm{CO(A}^1\Pi) + \mathrm{h}\nu_{\mathrm{An}}
\end{equation}

\begin{figure}[hbt]
\begin{center}
a)\includegraphics[clip = true,width=0.54\columnwidth]{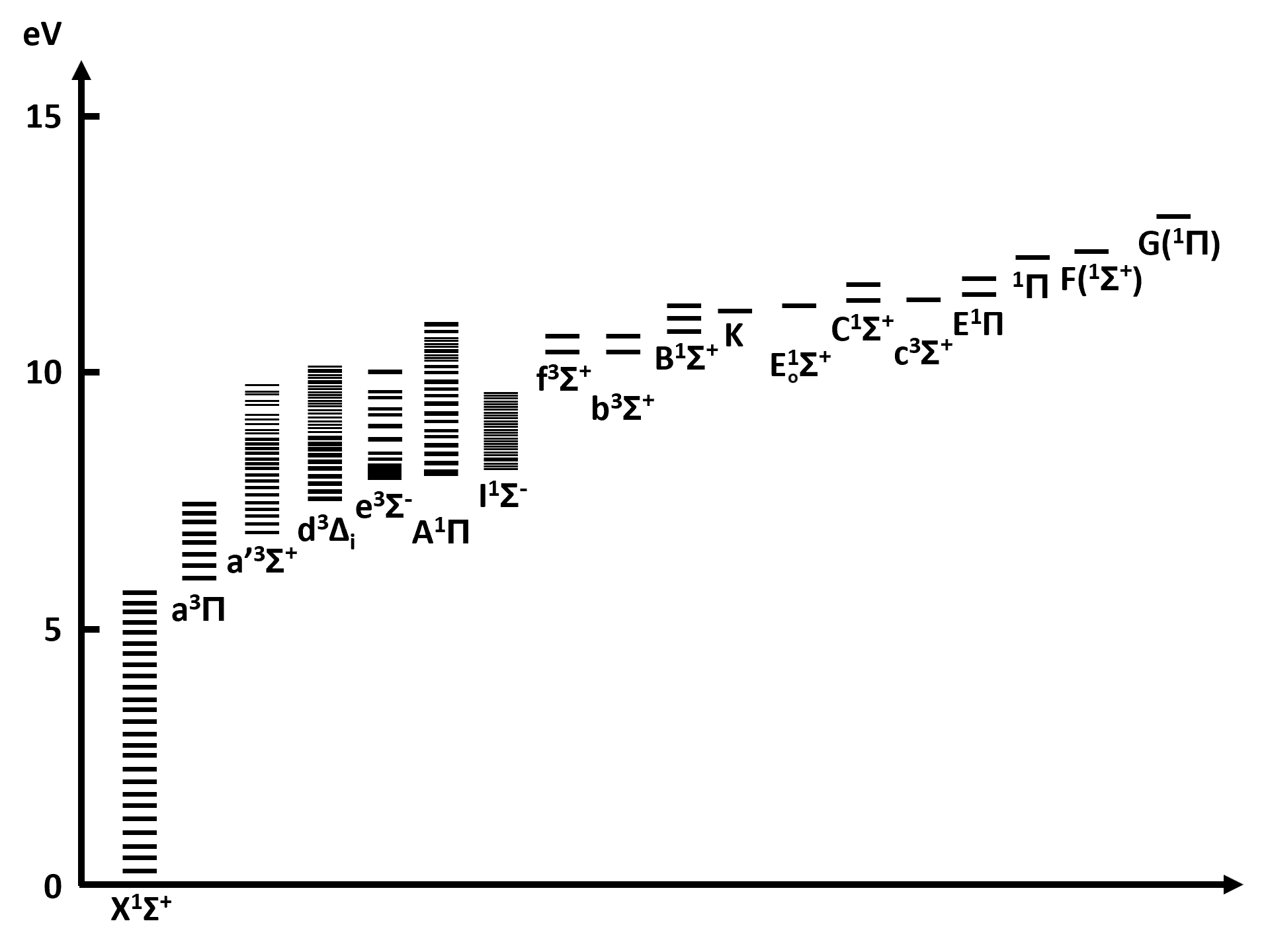}
b)\includegraphics[clip = true,width=0.32\columnwidth]{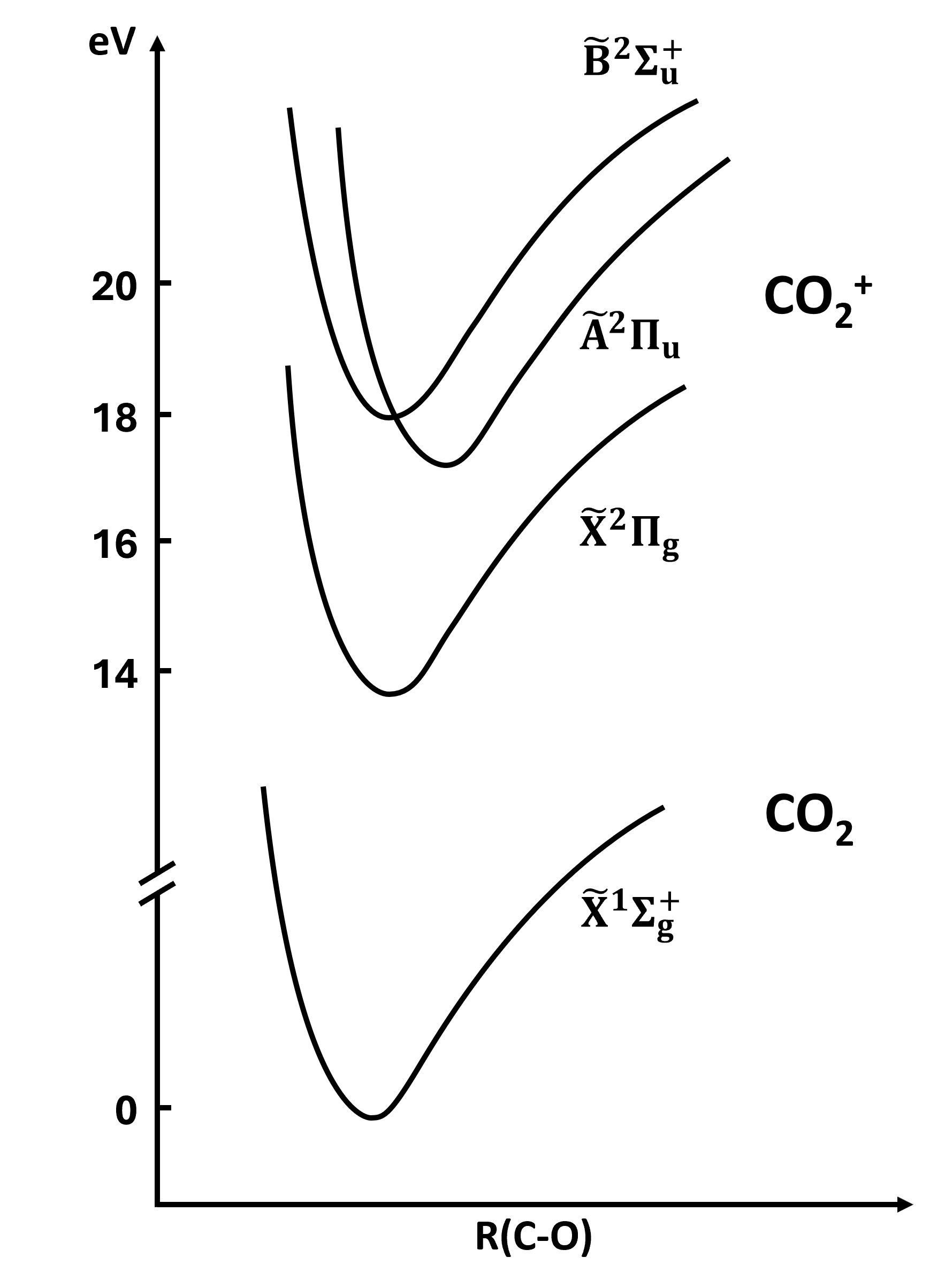}
\caption{Energy diagram for CO in part a), redrawn from \cite{krupenie1966}, and diagram of CO$_2$ and CO$_2^+$ in part b), redrawn from \cite{johnson1984}.}
\label{COenergy}
\end{center}
\end{figure}

\noindent
The most visible are the vibronic transitions from $\mathrm{CO(B}^1\Sigma^+, v' = 0)$ to $\mathrm{CO(A}^1\Pi, v'' = 0 - 5)$, ranging from 450 to approximatelly 650\,nm. At 440\,nm a weak emission from {\AA}ngstr{\"o}m system of vibronic transition (1$\rightarrow$1) can be expected, yet is indiscernible in the noise and possible overlap with other spectral bands, see Fig.\,\ref{comparison}a). 
The upper electronic state $\mathrm{CO(B}^1\Sigma^+, v' = 0)$ has excitation threshold energy of approx. 10.777\,eV \cite{itikawa2015}, compare the value in \cite{zawadzki2020,krupenie1966}. For better orientation, we have reproduced the energy diagram for CO in Fig.\,\ref{COenergy}a).

The spectrum of the multi-filament sinusoidal driven discharge  as well as of the APTD shows also intense bands of the 3$^{\mathrm{rd}}$ positive system of CO:

\begin{equation}
\label{3rd}
\mathrm{CO(b}^3\Sigma) \rightarrow \mathrm{CO(a}^3\Pi) + \mathrm{h}\nu_{\mathrm{3P}},
\end{equation}

\noindent
especially in the spectral interval below 340\,nm. The vibronic transitions (0$\rightarrow$0), (0$\rightarrow$1), (0$\rightarrow$2), (0$\rightarrow$3) and (0$\rightarrow$4) are visible. The upper electronic state $\mathrm{CO(b}^3\Sigma)$ has excitation threshold energy of approx. 10.399\,eV \cite{itikawa2015}, compare the value in \cite{zawadzki2020}. 

The bands of Asundi system are also visible in the range from 400 to 840\,nm (the identification according to \cite{krupenie1966, rond2008} is used). This system has a following denotation:

\begin{equation}
\label{asundi}
\mathrm{CO(a'}^3\Sigma^+) \rightarrow \mathrm{CO(a}^3\Pi) + \mathrm{h}\nu_{\mathrm{As}}.
\end{equation}

\noindent
Vibronic transitions (13$\rightarrow$2), (10$\rightarrow$1), (8$\rightarrow$0), (9$\rightarrow$1) and others were detected. The upper electronic state $\mathrm{CO(a'}^3\Sigma^+)$ has excitation threshold energy of approx. 6.863\,eV \cite{itikawa2015}, compare the values in  \cite{zawadzki2020,krupenie1966}.

The Triplet system originates from the radiative transition from state $\mathrm{CO(d}^3\Delta_i)$:

\begin{equation}
\label{triplet}
\mathrm{CO(d}^3\Delta_i) \rightarrow \mathrm{CO(a}^3\Pi) + \mathrm{h}\nu_{\mathrm{T}}
\end{equation}

\noindent
is also weakly visible. Especially the vibronic transitions (2$\rightarrow$0) and (1$\rightarrow$0) around 700\,nm. The upper electronic state $\mathrm{CO(d}^3\Delta_i)$ has excitation threshold energy of approx. 7.582\,eV \cite{krupenie1966}, compare the value in \cite{zawadzki2020}.

In some parts, possibly the weak spectra of Herzberg system may be detected as well with the dominant spectral bands from transitions (0$\rightarrow$1) and (0$\rightarrow$2) overlapping with Triplet, Asundi or other systems. The Herzberg system has a following denotation:

\begin{equation}
\label{Hz}
\mathrm{CO(C}^1\Sigma) \rightarrow \mathrm{CO(A}^1\Pi) + \mathrm{h}\nu_{\mathrm{Hz}}.
\end{equation}

\noindent
The upper electronic state $\mathrm{CO(C}^1\Sigma)$ has excitation threshold energy of approx. 11.396\,eV \cite{itikawa2015}, compare the value in  \cite{zawadzki2020}.

Besides the CO spectral bands, also the atomic lines of oxygen are visible for multi-filamentary and very weakly for the APTD discharge. Particularly, the atomic emission of the oxygen triplet at 777\,nm is clearly visible for multi-filamentary discharge:

\begin{equation}
\label{tripletO}
\mathrm{O(3p}\,^5\mathrm{P}) \rightarrow \mathrm{O(3s}\,^5\mathrm{S}) + \mathrm{h}\nu_{\mathrm{O777}}\mathrm{,}
\end{equation}

\noindent
with the upper state excitation energy of 10.74\,eV. It can be populated by electron impact dissociative excitation of oxygen molecule \cite{bajon2023,rond2008,mcconkey2008}. Similar excitation mechanism populates the upper state responsible for the atomic emission line at 843\,nm:

\begin{equation}
\label{tripletO}
\mathrm{O(3p}\,^3\mathrm{P}) \rightarrow \mathrm{O(3s}\,^3\mathrm{S}) + \mathrm{h}\nu_{\mathrm{O843}},
\end{equation}

\noindent
with the upper state excitation energy of 10.99\,eV.

The optical emission spectra in range 290 to 450\,nm also include spectral bands of CO$_2^+$, the upper state for Fox-Duffendack-Baker (FDB) system of the molecular ion has excitation threshold of approx. 17.3\,eV \cite{ajello1971} from the ground state of CO$_2$ molecule (note that the ionization of CO$_2$ by direct electron impact requires approx. 13.8\,eV \cite{itikawa2002}). Optical emission originates from following radiative process:

\begin{equation}
\label{fdb}
\mathrm{CO_2^+(A}^2\Pi_u) \rightarrow \mathrm{CO_2^+(X}^2\Pi_g) + \mathrm{h}\nu_{\mathrm{FDB}}
\end{equation}

\noindent
Additionally, the spectral bands of UV-doublet (also denoted as UV-$\lambda \lambda$ or $\lambda \lambda$2883-2896 system) can be detected, too, at wavelengths of approx. 288 and 289\,nm. 
The radiative transition reads:

\begin{equation}
\label{uv}
\mathrm{CO_2^+(B}^2\Sigma_u) \rightarrow \mathrm{CO_2^+(X}^2\Pi_g) + \mathrm{h}\nu_{\mathrm{UV}}.
\end{equation}

\noindent
This excited molecular ion $\mathrm{CO_2^+(B}^2\Sigma_u)$ has excitation threshold also of approx. 18.1\,eV \cite{ajello1971} (from ground state of CO$_2$ molecule). 
It is possible that other unidentified spectral bands or lines may be hidden in the noise or overlapped with the identified ones, we will however, focus our further analysis on the above mentioned ones.

From the identified OES in Fig.\,\ref{comparison} it is apparent that spectra of discharges driven between the plate electrodes (arrangement in Fig.\,\ref{chamber}c), i.e. the multi-filament and APTD, have rather similar appearance while the single-filament discharge (arrangement in Fig.\,\ref{chamber}b) is significantly different. 
We assign this effect to the conversion efficiency of the CO$_2$ to CO in the barrier discharge plasma, as the spectral bands for the single-filament arrangement are almost solely originating from emission of the CO$_2^+$ ion. No or very weak emission from CO molecules is observed in single-filament discharge. 
Indeed, for higher power or lower CO$_2$ flow and larger active plasma area the conversion to CO is more effective and it is visible in the optical emission spectra, compare \cite{kraus2002,brehmer2014,belov2016}. 

Some of the previous studies with low-scale barrier discharges in CO$_2$ show similar optical emission spectrum as for the single-filament setup here, reduced to spectral bands of CO$_2^+$. For example, Brehmer et al. \cite{brehmer2014} presented optical emission spectrum of barrier discharge in atmospheric pressure CO$_2$ in reactor with dimensions 70$\times$28$\times$1\,mm$^3$ (length along the gas flow, width and gap between the two quartz dielectrics). The spectrum is dominated by the bands of the UV-doublet (B$^2\Sigma_u^+ \rightarrow$ X$^2 \Pi_g$) and the Fox-Duffendack-Baker (A$^2\Pi_u \rightarrow$ X$^2 \Pi_g$) systems, both having the excitation threshold above 17\,eV. 
In that article, Brehmer et al. did not identify any other spectral systems.
Nevertheless, Cepelli et al. \cite{ceppelli2021} presented CO$_2$/CO spectra of a nanosecond pulse discharge (not a DBD) with different driving frequencies. It was shown that for lower pulse repetition rates and higher flow rate the CO bands vanished while the CO$_2^+$ bands dominated the spectrum. 
Navascu\'es et al. \cite{navascues2022} also reported the disappearance of the CO bands with decreasing of the applied voltage frequency. 
Note that the frequency is not responsible alone for the specific input energy (SIE) 
introduced to the gas and thus for the spectra, compare Fig.\,\ref{comparison}. 
The influence of the number of filaments (basically the SIE) on the CO$_2$ dissociation was investigated in \cite{douat2023}. 

\begin{figure}[hbt]
\begin{center}
\includegraphics[width=0.5\textwidth]{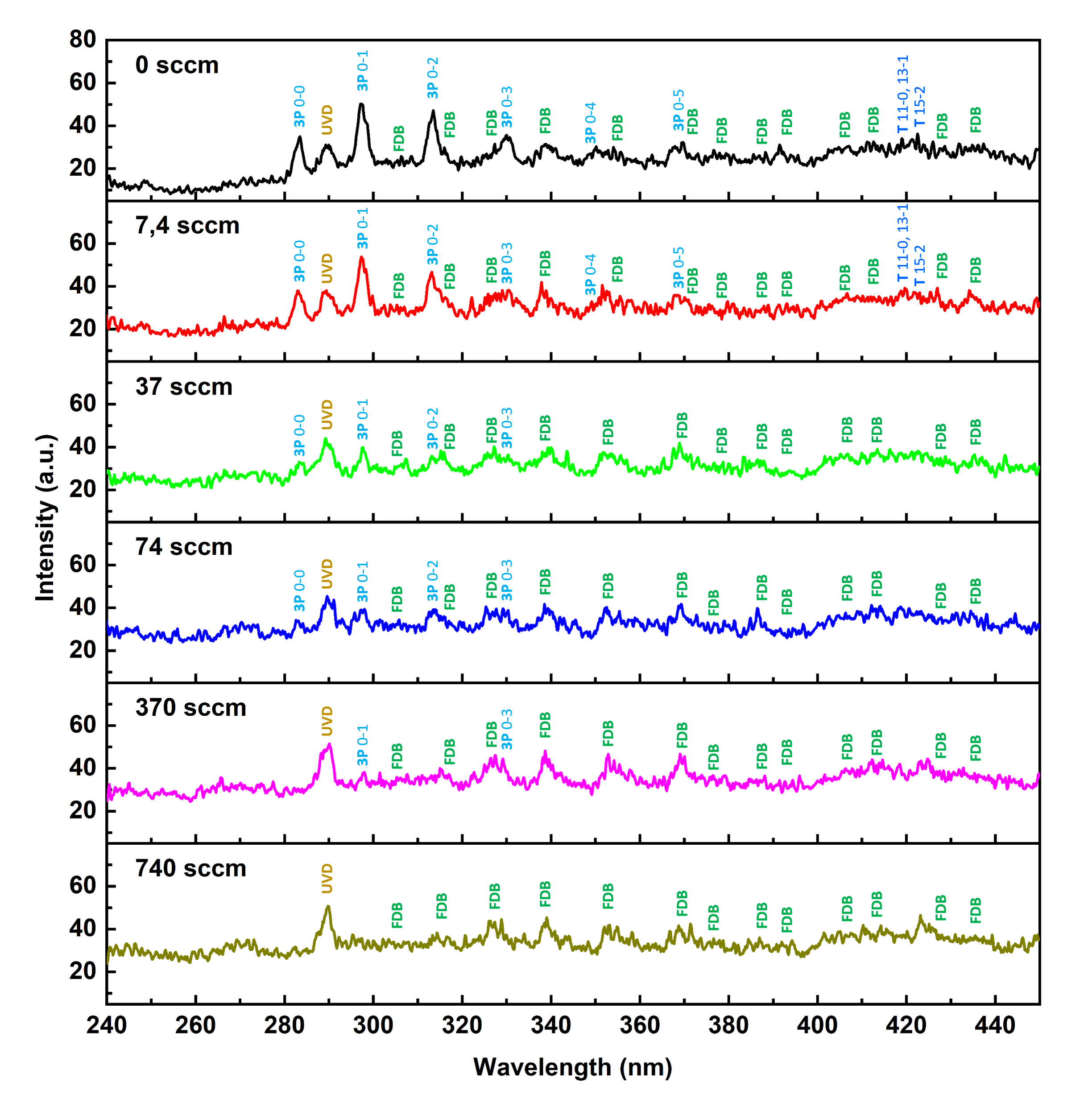}\\
\caption{Optical emission spectra for APTD discharge (2\,mm gap, 1\,kHz, 16\,kV$_{pp}$) taken at the same conditions as the FTIR spectra as presented in \cite{bajon2024} where the conversion $\alpha$ factor was measured.}
\label{spectra-alpha}
\end{center}
\end{figure}

We have performed our analysis of the SIE influence on the OES of an APTD and came to the same result, it is shown in Fig.\,\ref{spectra-alpha}. These spectra are measured for the same conditions as in \cite{bajon2024} where the CO$_2$ conversion was investigated, i.e. 2\,mm gap, 1\,kHz, 16\,kV$_{pp}$ for all presented flow rates (0 to 740\,sccm). 
With increasing flow rate, i.e. with decreasing SIE, the CO bands continuously diminish and finally disappear in the increasing CO$_2$ spectra.

In \cite{bajon2024} the conversion factor $\alpha = n_{CO}/(n_{CO}+n_{CO_2})$, obtained from the ground state densities, was experimentally determined using FTIR (Fourier-transform infrared) spectroscopy for three different flows: 7.4\,sccm results in $\alpha$ = 0.0043, 37\,sccm results in $\alpha$ = 0.00183 and 7.4\,sccm results in $\alpha$ = 0.00117. We use these results further in this paper for $E/N$ determination.

\section{\label{secX} Radiative states selection for $E/N$ determination and the collision-radiative model}

The intensity ratio method for reduced electric field $E/N$ determination is probably best known if used in nitrogen containing plasmas. Various plasmas were investigated using this method, from homogeneous APTD discharges \cite{bilek2019,mrkvickova2023}, through filamentary barrier discharges in volume or on surfaces \cite{jansky2021,jahanbakhsh2019correlation,liu2019,kosarev2012,stepanyan2014,bilek2023}, to positive streamers propagating between bare electrodes \cite{dijcks2022,shcherbakov2007,janda2017,brisset2019}. 
The method in nitrogen was developed and examined from multiple points of view and it proves to be useful in many cases. See for example in \cite{goldberg2022,mrkvickova2023,obrusnik2018electric,pancheshnyi2006,hoder2016,paris2005intensity,paris2006,malagon2019}, to name a few. Here, we will use this understanding and attempt to develop similar method also for the discharges in CO$_2$.

The intensity ratio method compares 
emitted intensities of optical transitions coming from two radiative states
with different excitation energy threshold. Both radiative states are supposed to be populated by direct electron impact from the same state, typically the ground state of the atom/molecule of the gas within which the plasma is operated. If the original ground states are different, their relative densities need to be known (the following equations will clarify this). Due to the fact that each of the two radiative states has a different excitation energy threshold and/or the cross-sections for electron excitation are of different form, each of the two states is populated with different rate for a given local mean electron energy. The mean electron energy is an unambiguous function of the local electric field $E$ for given gas density $N$. The electron energy distribution function has to be relaxed. The ratio of the two electron excitation rate coefficients, which are dependent solely on the reduced electric field $E/N$, gives the information about the local reduced electric field. 
Let's consider the simplest rate  equation for density of the radiative state $n_1$ at given spatial coordinate:

\begin{equation}
\label{population1}
\frac{\mathrm{d}n_{1}(t)}{\mathrm{d}t} = n_e(t) \cdot n_{g} \cdot k_{g,1}(E(t)/N) - \frac{n_{1}(t)}{\tau_{\mathrm{eff,1}}(t)}  
\end{equation}

\noindent
where the temporal change of the radiative state density is determined by three processes: population of the state due to the direct electron impact from the ground state which is given by the first term on the right side of the equation, depopulation due to the radiative and collisional quenching, as it is given by effective lifetime in the second term on the right side of the equation. 
These processes quantified in equation (\ref{population1}) represent the simple collision-radiative model (CRM). 
$n_1(t)$ is the state 1 number density, $n_e(t)$ is the electron number density, $n_g$ is the number density of the ground state, $k_{g,1}(E(t)/N)$ is the rate coefficient as a function of $E/N$ and $\tau_{\mathrm{eff,1}}(t)$ is the effective lifetime which may be time-dependent in general, we will assume it to be time-independent for our case as a reasonable assumption. It is defined as:

\begin{equation}
\label{effective}
\frac{1}{\tau_{\mathrm{eff,1}}} = k_{\mathrm{CO_2,1}}n_{g} + \sum_{\nu''=0}^{\infty} \frac{1}{\tau_{\nu'\nu'',1}}
\end{equation}

\noindent
where $k_{CO_2,1}$ is rate coefficient for collisional quenching by the surrounding gas molecules, in our case CO$_2$ and $\tau_{\nu'\nu'',1}$ are radiative lifetimes for the state deexcitation to vibrational states of lower electronic state, as a sum $\frac{1}{\tau_{rad,1}} =  \sum_{\nu''=0}^{\infty} \frac{1}{\tau_{\nu'\nu'',1}} = A_1$. Where $A_1$ is the Einstein coefficient of spontaneous emission resulting in all spectral system 
bands.

Rewriting the above mentioned equations also for a second state of the same molecule (state 2 with the density of $n_2$), say with higher threshold energy, we can divide these two equations and after some modifications we obtain:

\begin{equation}
\label{population2}
\frac{
\frac{\mathrm{d}n_{2}(t)}{\mathrm{d}t} 
+ \frac{n_{2}(t)}{\tau_{\mathrm{eff,2}}}
}{
\frac{\mathrm{d}n_{1}(t)}{\mathrm{d}t} 
+ \frac{n_{1}(t)}{\tau_{\mathrm{eff,1}}}
}
= \frac{ k_{g,2}(E(t)/N)}{k_{g,1}(E(t)/N)} = \frac{ k_{g,2}}{k_{g,1}}(E(t)/N)
\end{equation}

\noindent
Note that the electron and ground state densities are gone. We can rewrite the equation using the formula for intensity of the optical emission $I = TnhcA_{\nu'\nu''}/\lambda$ where $n$ is the number density of the radiative state and $T$ is a transmission coefficient of the apparatus. As all optical spectra measurements were corrected on relative intensities using calibration lamp, the $T$ is equal unity here. $h$ is Planck constant, $c$ speed of light, $A_{\nu'\nu''}$ is the rate of spontaneous emission for given vibronic transition $\nu'\nu''$ (band Einstein coefficient) and $\lambda$ the emission wavelength. We receive:

\begin{eqnarray}
\label{master}
\frac{
\frac{\mathrm{d}I_{2}(t)}{\mathrm{d}t} 
+ \frac{I_{2}(t)}{\tau_{\mathrm{eff,2}}}
}{
\frac{\mathrm{d}I_{1}(t)}{\mathrm{d}t} 
+ \frac{I_{1}(t)}{\tau_{\mathrm{eff,1}}}
}
&= \frac{ k_{g,2}(E(t)/N)T_2A_{\nu'\nu'',2}\lambda_1}{k_{g,1}(E(t)/N)T_1A_{\nu'\nu'',1}\lambda_2} \nonumber\\
&= R_{t-d}(E(t)/N).
\end{eqnarray}

\noindent
If the collisional quenching and radiative lifetimes data are known from literature the effective lifetimes as well as the Einstein coefficients are given. Thus, on the left side of the equation are the measured intensities as an output from the experiment and on the right side the function of the reduced electric field, sometimes called as calibration curve (for given time instant). 
The rate coefficients, and thus the $R(E/N)$ function, can be obtained by solving the Boltzmann equation for selected set of cross-sections, using e.g. the project LXCat \cite{pitchford2017}. 
The above mentioned equation (\ref{master}) is a time-dependent model and is usable also for a non-steady-state regimes in plasmas, even for such strongly non-steady-state processes as ionization waves or streamer discharges. In that case, the recording device with high temporal resolution is necessary.

If we investigate discharge in a steady-state regime, as for example the APTD, the equation\,(\ref{master}) loses the time derivatives (as they are negligible) and results in:

\begin{equation}
\label{mastersteady}
\frac{
I_{2}
}{
I_{1}
}
= \frac{ k_{g,2}(E/N)T_2A_{\nu'\nu'',2}\lambda_1}{k_{g,1}(E/N)T_1A_{\nu'\nu'',1}\lambda_2} \cdot \frac{\tau_{\mathrm{eff,2}}}{\tau_{\mathrm{eff,1}}}
= R_{s-s}(E/N).
\end{equation}

\noindent
Note that the ratio of effective lifetimes moved to the right side of the equation to isolate the measured intensities on the left which makes the equation practical. This is an equation of a steady-state model. It is a special case of the time-dependent approach.

In the case when the electron impact excitation takes place from different ground state molecules, e.g. CO$_2$ and CO for the intensity ratio utilization in CO$_2$/CO mixtures, the relative number densities has to be known. For such case one can use the FTIR measurements and determine the conversion factor $\alpha = n_{CO}/(n_{CO}+n_{CO_2})$. We can then write the equation\,(\ref{mastersteady}) (which now includes the number densities of CO and CO$_2$) in following way:

\begin{eqnarray}
\label{mastersteady2}
\frac{I_{2}}{I_{1}} 
&= \frac{ k_{g,2}(E/N)n_{CO_2}T_2A_{\nu'\nu'',2}\lambda_1}{k_{g,1}(E/N)n_{CO}T_1A_{\nu'\nu'',1}\lambda_2} \cdot \frac{\tau_{\mathrm{eff,2}}}{\tau_{\mathrm{eff,1}}} \nonumber\\
&= R_{s-s}(E/N)\frac{n_{CO_2}}{n_{CO}} \nonumber\\
&= R_{s-s}(E/N)\frac{1-\alpha}{\alpha}.
\end{eqnarray}

The above described simplified models are considered appropriate for single shot or low frequency repetitive plasmas or discharges, with low pre-ionization and negligible densities of excited species (like metastables) prior to every new plasma occurrence. This is valid for discharges in electronegative gases and close-atmospheric pressure plasmas with effective collisional quenching. The above mentioned conditions were verified for different discharges in air and here we undertake the first  application of such simplified model to CO$_2$ plasmas. 
Taking into account the successful verification of the necessary conditions for air (see in \cite{obrusnik2018electric} and other references as given in the beginning of this section), we are of the opinion that the use of the method for CO$_2$ is also appropriate (we will discuss the necessary next steps at the end of the article). The missing point is to select the suitable radiative states for the use of the method. 

In the previous section, the OES measurements showed that for different arrangements, flows or discharge power the concentration of CO in the discharge area may differ and most probably may be varying during the discharge operation. 
For the method development it is therefore reasonable, and the most convenient solution, to select two radiative states which are both populated from the same ground state, either from the CO or from the CO$_2$ molecule. Nevertheless, for experiment with APTD where the conversion factor was measured (and the concentration of CO and CO$_2$ molecules is thus known) we discuss the method given by the equation (\ref{mastersteady2}) as well, later in this article.

First we will consider the spectra for the use in single-filament barrier discharge, originating from the electron impact ionizing excitation of the CO$_2$ molecule. 
The only solution is the use of the two spectral bands of CO$_2^+$, the FDB and the UV-doublet, as no other spectra originating from different radiative states are easily detectable. 
The excitation energy thresholds of their radiative states are nevertheless close, for $\mathrm{CO_2^+(A}^2\Pi_u)$ of the FDB it is 17.3\,eV and for $\mathrm{CO_2^+(B}^2\Sigma_u)$ of the UV-doublet 18.1\,eV. This fact may certainly be a source of increased uncertainty of the method based on these two states, yet even such states should make the method work. The sensitivity of ratio of optical emission intensities coming from radiative states of even closer excitation energy thresholds on the electric field was shown in \cite{kusyn2023} and its theoretical application was inspected in \cite{kusyn2024}.

\begin{figure}[hbt]
\begin{center}
\includegraphics[width=0.5\textwidth]{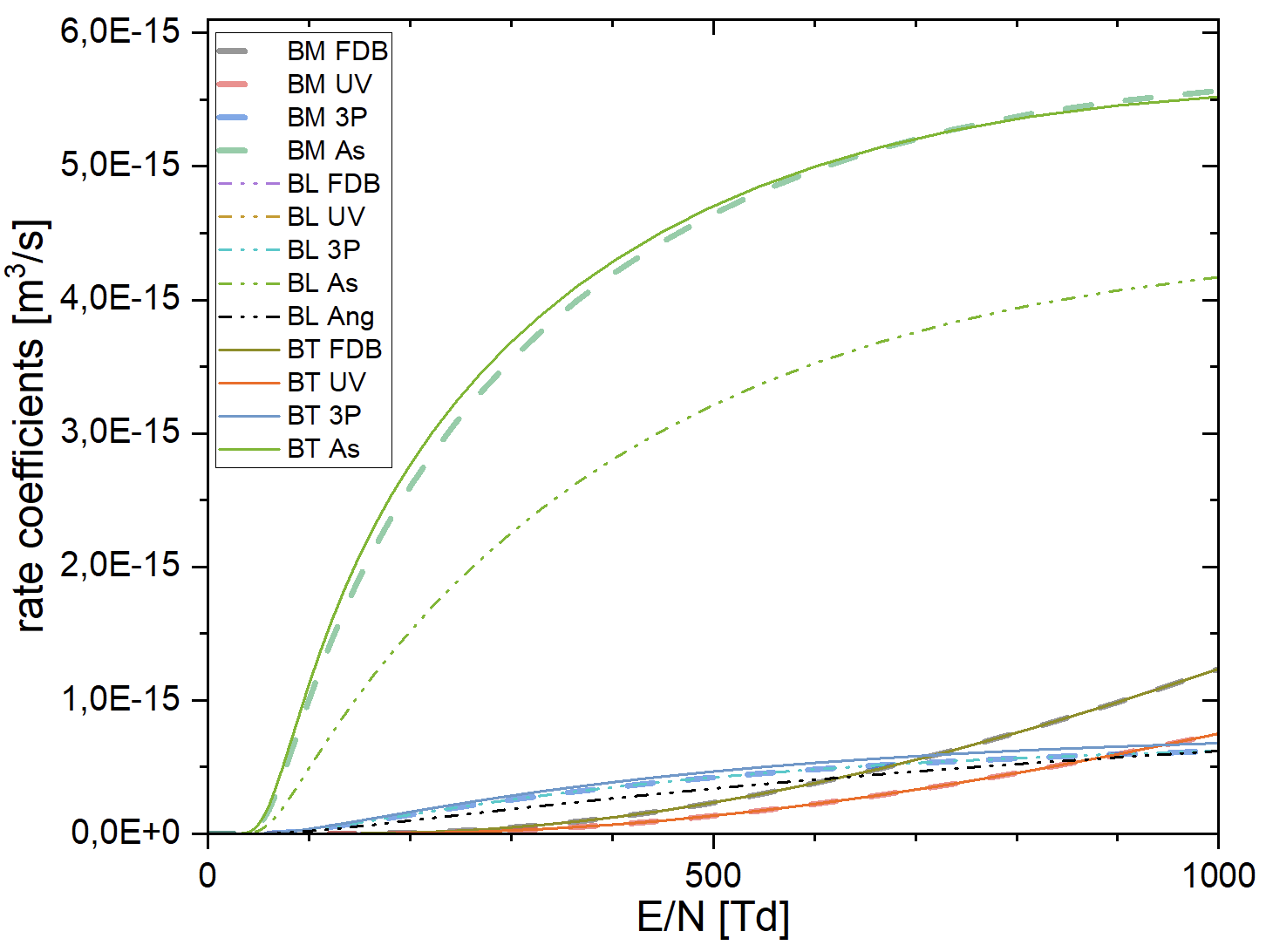}
\caption{Rate coefficients for populating electronic radiative states of FDB and UV-doublet systems from CO$_2$ ground state and radiative states of 3$^{\mathrm{rd}}$ positive (3P), Asundi (As) and {\AA}ngström (Ang) systems from CO ground state. The excitation Franck-Condon factors are therefore not included here. The first two letters in legend denote the used database.}
\label{rates}
\end{center}
\end{figure}

The rate coefficients for the population of the radiative states of $\mathrm{CO_2^+(A}^2\Pi_u)$ and $\mathrm{CO_2^+(B}^2\Sigma_u)$ as computed by BOLSIG+ \cite{hagelaar2005} using Biagi database for CO$_2$ \cite{biagi} taken from LXCat are shown in Fig.\,\ref{rates}. The mixture used through this article for the LXCat computations was always 99\% of CO$_2$ and 1\% of CO at 300\,K. The slight variations of rate coefficients due to the CO creation may be neglected \cite{bajon2023}. 

To be complete, in Fig.\,\ref{rates}, the first two  letters in the legend denote the used database: BM stands for Biagi (CO$_2$) and Morgan (CO), BL for Biagi (CO$_2$) and Lisbon (CO) and BT for Biagi (CO$_2$) and Triniti (CO). The Biagi input is present for all combination as it includes the excitation cross-sections for $\mathrm{CO_2^+(A}^2\Pi_u)$ and $\mathrm{CO_2^+(B}^2\Sigma_u)$. The cross-sections for CO may differ. The cross-sections for 3$^{\mathrm{rd}}$ positive (b$^3\Sigma^+ \rightarrow$ a$^3\Pi$) and Asundi (a'$^3\Sigma ^+ \rightarrow$ a$^3\Pi$) systems radiative states b$^3\Sigma^+$ and a'$^3\Sigma ^+$ are different for each database, i.e.: the Lisbon database uses the cross-sections of Sawada et al. \cite{sawada1972} for both CO radiative states, the Morgan database uses Sawada et al. cross-section for b$^3\Sigma^+$ excitation and Lin et al. (reference not specified in LXCat) for a'$^3\Sigma ^+$ state excitation, the Triniti database uses for both states cross-sections from Land et al. \cite{land1978}. We did not used the data from Zobel et al. \cite{zobel1996}, as proposed in \cite{itikawa2015}, yet. 
The Lisbon set for CO was the only containing the cross-sections for the {\AA}ngström system radiative state \cite{kanik1993,itikawa2015}. 
A detailed  uncertainty quantification and sensitivity analysis (see e.g. Obrusník et al. \cite{obrusnik2018electric} and Bílek et al. \cite{bilek2018electric}) of the method, taking into account critical revision of accessible data in the literature, is planned for the future research. 

The considered cross-sections for the radiative states excitations are for the electronic states, i.e. integral cross-sections for all vibrational states included. 
For excitation of selected upper vibrational states (responsible for the whole spectral band emission due to the $\nu'\nu''$ transition) it is necessary therefore to include also the excitation Franck-Condon factors. 
Additionally, the collisional quenching rates and radiative lifetimes need to be found in literature as well, in order to have a full set of parameters for the equation (\ref{master}). 
For further analysis, we have selected following radiative transition of CO$_2^+$ available in the OES: $A^2\Pi_{u_{3/2}}(\nu'=1) \rightarrow X^2\Pi_{g_{3/2}}(\nu''=0)$ with resulting spectral band with head at 337.0\,nm. 
The list of found values for the mentioned parameters is summarized in table \ref{tableCO2A}. Note that the vibronic state of CO$_2$ molecule is described by three vibrational numbers, e.g. (0,0,0), yet the vibronic transition, change of the first vibrational number, from the 1 to 0 is described as (1$\rightarrow$0). 
Similarly for CO, the transitions are also denoted by an arrow.

\begin{table*}[htb]
\center
\caption{Excitation Franck-Condon factors, radiative lifetimes $\tau_{rad}$ and collisional quenching rates $k_{CO_2}$ for radiative state $\mathrm{CO_2^+(A}^2\Pi_{u_{3/2}})$, responsible for Fox-Duffendack-Baker spectral system, with given wavelengths of spectral band heads for given vibrational transition. The Einstein coefficients $A_{\nu'\nu''}$ for vibronic transitions were obtained using excitation Franck-Condon factors, radiative lifetimes and emission probabilities.
}
\label{tableCO2A}
\begin{tabular}{lc}\br
Radiative state:  & $A^2\Pi_{u_{3/2}}(\nu'=1)$  \\\br
Band head [nm]:   & 337.0 (1$\rightarrow$0)$_{3/2}$ \\\mr
Exc. FC factor:    & 0.21 \cite{nishimura1968}     \\
   & 0.18 \cite{farley1996}  \\
   & 0.26 \cite{tokue1990} \\\mr
 $\tau_{rad}$ [ns]:  
 &   109$\pm$11 \cite{hesser1968} \\
  &  100$\pm$8 \cite{schlag1977}\\
 &   122$\pm$6 \cite{maier1980}\\\mr
 $A_{\nu'\nu''}$ [s$^{-1}]$:   & 1.862$\times10^{6}$ \cite{nishimura1968,ajello1971,hesser1968}  Nishimura et al.      \\
  & 2.151$\times10^{6}$ 
  \cite{tokue1990,ajello1971,hesser1968} Farley et al. \\\mr
Total $k_{CO_2}$ [cm$^{3}$s$^{-1}$]: & 10.5$\times10^{-10}$ \cite{farley1996}, 8.6$\times10^{-10}$ \cite{alderson1973}, \\
& 2$\times10^{-10}$ \cite{mackay1972}, 1.8$\times10^{-10}$ \cite{farley1997} \\\br
\end{tabular}
\end{table*}

The parameters selected for the radiative state $\mathrm{CO_2^+(A}^2\Pi_u)$ were considered in a following way. As excitation Franck-Condon factors, two sets were used to set the borderlines of uncertainty interval possibly caused by their selection. First set we denote by `N' as it is based on the work of Nishimura \cite{nishimura1968}, i.e.: 0.21 for the excitation to $\mathrm{CO_2^+(A}^2\Pi_{u_{3/2}})(\nu'=1)$ from the CO$_2$ ground state. This radiative state is responsible for the FDB band emission with head at 337\,nm. 
The band selection was given by considering the overlap of the close spectral bands and due to the linear dispersion of the setup. 
The second set we denote as `TF' as it is based on the articles of Tokue et al. \cite{tokue1990} and Farley et al. \cite{farley1996}, i.e.: 0.18 for the excitation to $\mathrm{CO_2^+(A}^2\Pi_{u_{3/2}})(\nu'=1)$ from the CO$_2$ ground state.

As radiative lifetimes $1/A$ we used the selective (vibrational state dependent) values from Hesser et al. \cite{hesser1968}, i.e.: 109$\pm$11\,ns for the (1$\rightarrow$0)$_{3/2}$ transition. 
Note that the small deviations from these values, as found in literature, are usually within the error bars and the resulting $E/N$ are not that sensitive to this particular parameter.

From the quenching rates values for the $\mathrm{CO_2^+(A}^2\Pi_u)$ radiative states given in the table\,\ref{tableCO2A}, we used in further analysis the value 8.6$\times10^{-10}$\,cm$^{3}$s$^{-1}$ from Alderson et al. \cite{alderson1973} as it was measured at the highest pressure of 100\,Torr, i.e. the closest to the atmospheric pressure conditions. The remaining values were obtained typically in mTorr range. 
 The resulting effective lifetime used for the $\mathrm{CO_2^+(A}^2\Pi_u)$ radiative states is $\tau_{eff}^{FDB} = 48$\,ps for 300\,K.

The parameters for the radiative state $\mathrm{CO_2^+(B}^2\Sigma_u^+)$ are given in table\,\ref{tableCO2B}. The radiative transition for UV-doublet as given in eq.\,(\ref{uv}) can be specified using following two notations where each transition is responsible for one of the two dominant spectral bands: $\mathrm{CO_2^+(B}^2\Sigma_u)(0,0,0) \rightarrow \mathrm{CO_2^+(X}^2\Pi_{g_{3/2}})(0,0,0) + \mathrm{h}\nu_{\mathrm{UV}}$ for 288.2\,nm and $\mathrm{CO_2^+(B}^2\Sigma_u)(0,0,0) \rightarrow \mathrm{CO_2^+(X}^2\Pi_{g_{1/2}})(0,0,0) + \mathrm{h}\nu_{\mathrm{UV}}$ for 289.6\,nm, according to \cite{ajello1971}. The radiative deexcitation of the radiative state via the band at 288.2\,nm is approx. 49\% while of the band at 289.6\,nm it is 51\% \cite{ajello1971}. 
Detailed rovibronic structure of the transition in eq.\,(\ref{uv}) can be found in \cite{johnson1984}. For $1/A$ we use the value 118$\pm$12\,ns \cite{hesser1968,mackay1972}. For the collisional quenching rate coefficient the value of 7.5$\times10^{-10}$ \cite{alderson1973} is taken from Alderson et al., too, for the same reason as above discussed value for $\mathrm{CO_2^+(A}^2\Pi_u)$ state. The resulting effective lifetime used for the $\mathrm{CO_2^+(B}^2\Sigma_u^+)$ radiative state is $\tau_{eff}^{UV} = 55$\,ps for 300\,K.

\begin{table*}[htb]
\center
\caption{Excitation Franck-Condon factors, radiative lifetimes $\tau_{rad}$ and collisional quenching rates $k_{CO_2}$ for radiative state $\mathrm{CO_2^+(B}^2\Sigma_u^+)$, responsible for the UV-doublet radiation, with given wavelengths of the doublets spectral band peaks  for given vibrational transition. 
}
\label{tableCO2B}
\begin{tabular}{lc}\br
Radiative state: & $\mathrm{CO_2^+(B}^2\Sigma_u^+)$ (0,0,0)  \\\br
Band head [nm]: & dominant bands at 288.3 and 289.7\,nm (0$\rightarrow$0)   \\\mr
Exc. FC factor: & radiative state identical for both spectral bands 
\\\mr
$\tau_{rad}$ [ns]:  &  135$\pm$10 \cite{alderson1973}, 118$\pm$12 \cite{hesser1968}, 140$\pm$7 \cite{maier1980}, \\
& 140 \cite{freeman1974}, 118$\pm$12 \cite{mackay1972}, 117.4$\pm$2.1 \cite{smith1975}\\\mr
$A_{\nu'\nu''}$ [s$^{-1}]$: & 8.476$\times10^{6}$ \cite{ajello1971,hesser1968} \\\mr 
$k_{CO_2}$ [cm$^{3}$s$^{-1}$]: &  6.75$\times10^{-10}$ \cite{farley1996}, 7.5$\times10^{-10}$ \cite{alderson1973},  19$\times10^{-10}$ \cite{freeman1974},     \\
&  11$\times10^{-10}$ \cite{mackay1972}, 3.1$\times10^{-10}$ \cite{farley1997} \\\br
\end{tabular}
\end{table*}

The case for the optical emission spectra from CO radiative states can be more problematic. As the excitation to the CO radiative states may not originate solely from direct electron impact to CO molecule at the ground state, but also via CO$_2$ molecule splitting via interaction with faster electrons. 
Here, we focus our analysis on the first option, see the explanation in the following text, and we leave the second for future research. The intensity ratio of CO spectral bands is used only in the case of APTD discharge in CO$_2$ later in this article. Bajon et al. \cite{bajon2023} have shown that the reduced electric field in the APTD peaks at approx. 145\,Td. With relaxed EEDF under such conditions, only approximatelly 1\% of electrons reaches the energy of 12 or 13\,eV. 
The dominant source of CO in CO$_2$ discharges is via CO$_2$ dissociation by electron impact, with thresholds of 7 and 11\,eV for resulting CO ground and excited $\mathrm{CO(a}^3\Pi)$ states, respectively. 
If  the intensity ratio is constructed using the states lying higher than $\mathrm{CO(a}^3\Pi)$ state (additional 1 and 4\,eV higher, see further in the text), the contribution of the CO$_2$ dissociation to production of  higher radiative states of CO in APTD may be therefore neglected. 
The stepwise excitation is at atmospheric pressure CO$_2$ negligible as well due to effective collisional quenching, see further in the text.

As mentioned above, for the intensity ratio method in atmosphere with CO gas component (for APTD), we propose to use the emission from radiative state $\mathrm{CO(b}^3\Sigma)$ 
which is responsible for the optical emission as spectral system of the 3$^{\mathrm{rd}}$ positive and above mentioned state $\mathrm{CO_2^+(B}^2\Sigma_u^+)$ of UV-doublet. 
The main reason is that the difference of the excitation energy thresholds for the respective radiative states is relatively big. 
Based on the overview spectra, we have selected the (0$\rightarrow$0) band of 3$^{\mathrm{rd}}$ positive 
due to its isolated position without overlap with neighboring bands. 
The parameters for the radiative state $\mathrm{CO(b}^3\Sigma)$ is given in table\,\ref{tableCO}.

\begin{table}[htb]
\center
\caption{Excitation Franck-Condon factors, radiative lifetimes $\tau_{rad}$ and collisional quenching rates $k_{CO_2}$ for radiative state $\mathrm{CO(b}^3\Sigma)(\nu' = 0)$, responsible for the 3$^{\mathrm{rd}}$ positive system radiation, with given wavelength of the spectral band head for given vibrational transition. 
}
\label{tableCO}
\begin{tabular}{lc}\br
Radiative state: & $\mathrm{CO(b}^3\Sigma)(\nu' = 0)$ \\\br
Band head [nm]: & 283\,nm (0$\rightarrow$0)   \\\mr
Exc. FC factor: & 0.9690 \cite{zawadzki2020}  \\\mr
$\tau_{rad}$ [ns]:  &  53.6$\pm$0.3 \cite{smith1973}  \\
&  57.6$\pm$1.24 \cite{rogers1970}  \\
&  56$\pm$1 \cite{sprang1977}  \\
 &  57.8$\pm$0.3 \cite{qin2017}   \\\mr
$A_{\nu'\nu''}$ [s$^{-1}$]:  &  2.625 $\times10^{6}$ \cite{silva2006} \\\mr
$k_{CO}$ [cm$^{3}$s$^{-1}$]: & 2.8875$\times10^{-10}$ \cite{twist1979} \\
& 7.5$\times10^{-10}$ \cite{smith1973} \\\br
\end{tabular}
\end{table}

The variation of radiative lifetime values for $\mathrm{CO(b}^3\Sigma)(\nu' = 0)$ does not have any significant influence under given conditions, the values are all close to each other. 
We have used the value of $1/A^{3rd} = 57.6$\,ns. 
Note that no data were found on the quenching of $\mathrm{CO(b}^3\Sigma)(\nu' = 0)$ state by CO$_2$, only via CO molecules which we used. 
Even for quenching by CO, there are values differing by a factor up to 2.5, see table\,\ref{tableCO} and \cite{rogers1970}. We have used the value of Twist {\it et al.} \cite{twist1979}, see the table. 
This fact inserts another uncertainty into our results and should be corrected for further uses by additional analysis or direct measurement of these coefficients. 
The resulting effective lifetimes are for $\mathrm{CO(b}^3\Sigma)(\nu' = 0)$ $\tau_{eff}^{3rd} = 143$\,ps for 300\,K.

The last possible component for the ratio method utilization based on the observed spectra is the emission from the {\AA}ngström system, which we leave for future work. 
We also did not use the Asundi bands, due to differing excitation cross-sections (see Fig.\ref{rates} and discussion in the text), very long radiative lifetime and missing CO$_2$ quenching rates in the literature.

\begin{figure}[hbt]
\begin{center}
\includegraphics[width=0.44\textwidth]{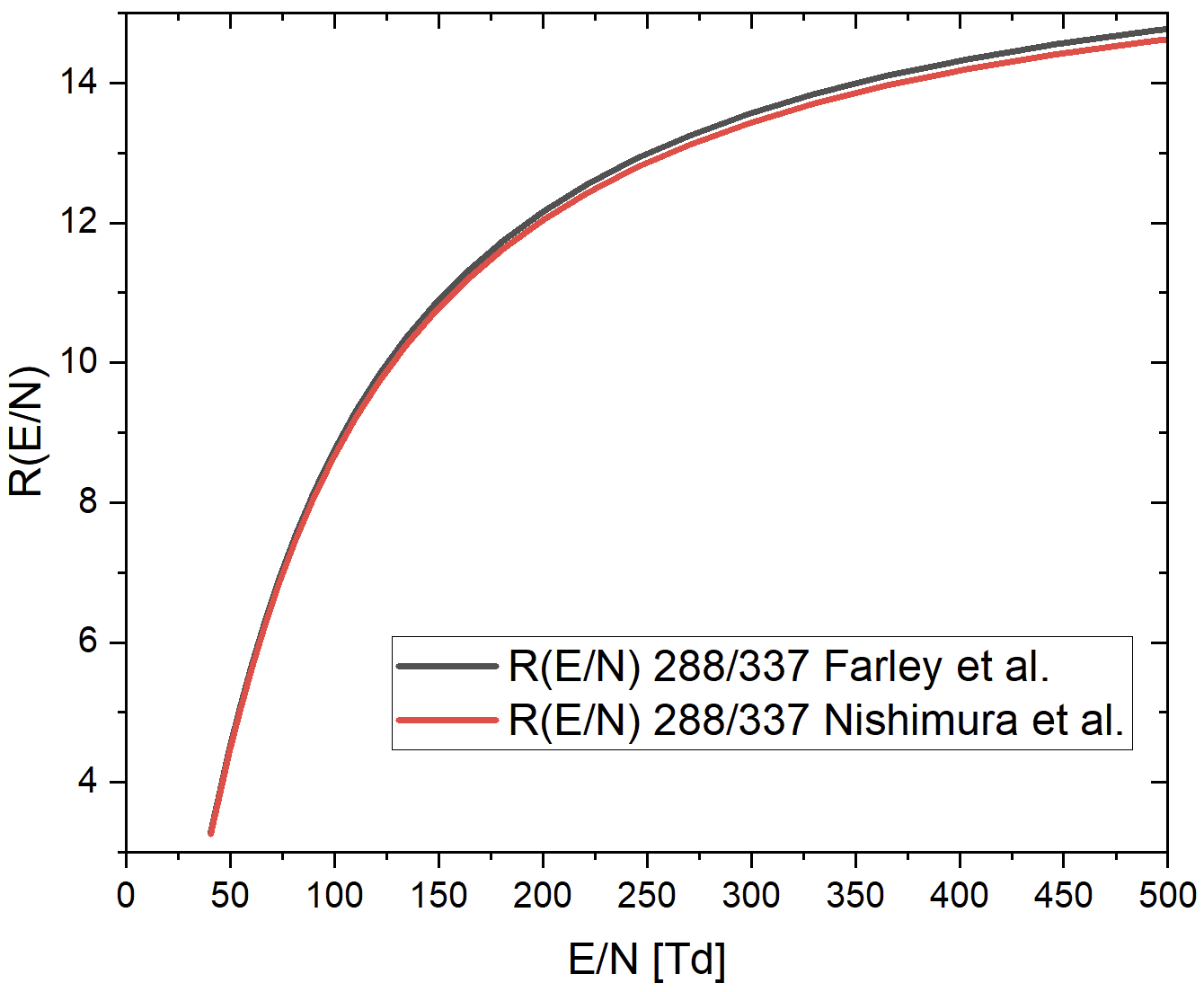}
\caption{The $R(E/N)$ function CO$_2^+$ spectral bands intensity ratio. 
The function $R_{t-d}(E/N)$ (equation\,\ref{master}) for the ratio of UV-doublet (0$\rightarrow$0) and Fox-Duffendack-Baker (1$\rightarrow$0)$_{3/2}$ system bands intensities are presented for two sets of excitation Franck-Condon factor values, all with almost identical rate coefficients for BM, BL or BT databases.}
\label{ratioss}
\end{center}
\end{figure}

In Fig.\,\ref{ratioss}, the ratios $R_{t-d}(E/N)$ (equation\,(\ref{master})) are shown as a functions of $E/N$, for radiative states of 
UV-doublet/FDB systems for the different excitation Franck-Condon factor sets, from Nishimura and Tokue and Farley, as discussed earlier. They do not differ significantly.  
This ratio was used for the single filament experiment, and after correction, also for the APTD, see further.

\begin{figure}[hbt]
\begin{center}
\includegraphics[width=0.5\textwidth]{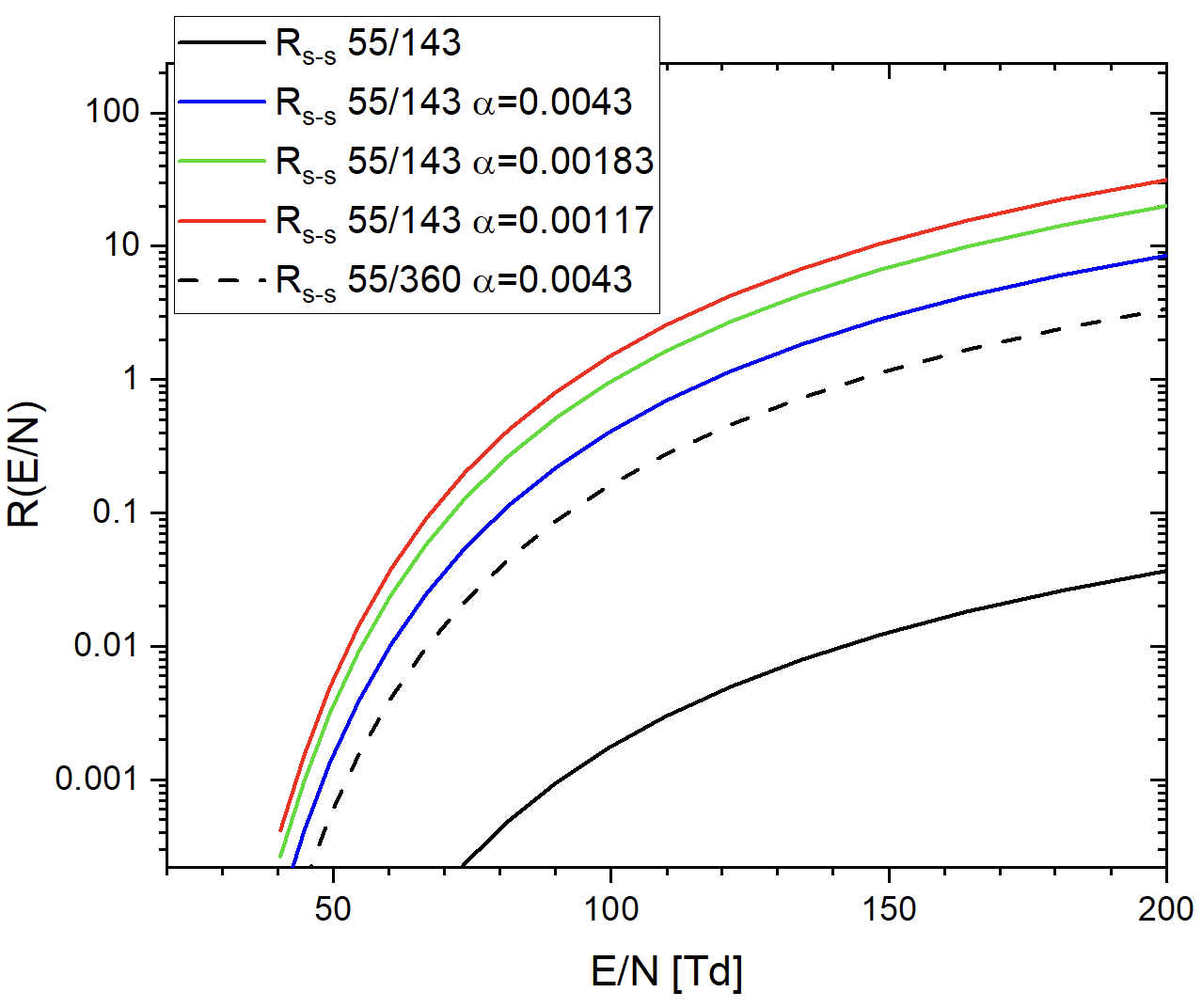}
\caption{The $R_{s-s}$ ratio function for the radiative states of UV doublet/3$^{\mathrm{rd}}$ positive systems with different quenching rate sets of the 3$^{\mathrm{rd}}$ positive system, once with CO quenching and effective lifetime of 143\,ps (full lines) and with best fitting quenching value of 360\,ps for $E/N$ = 125\,Td for $\alpha$ = 0.0043 with ratio value of 0.49 in APTD (dashed line).}
\label{ratiossss}
\end{center}
\end{figure}

The other investigated ratio was using the combined CO$_2^+$ and CO emissions. Particularly the emission of the UVD band at 288\,nm and the 3$^{\mathrm{rd}}$ positive band at 283\,nm. The $R_{s-s}$ ratio functions are shown in Fig.\,\ref{ratiossss} for various conditions. 
As the radiative states originate from electron impact excitation of different ground states, the equation\,(\ref{mastersteady2}) was used together with the known conversion factors. 
The full-lines represents the ratio function using CO$_2$ quenching for UVD ($\tau_{eff}^{UV}$ = 55\,ps) and CO quenching for 3$^{\mathrm{rd}}$ positive system ($\tau_{eff}^{3rd}$ = 143\,ps). 
The ratio function $R_{s-s}$ resulting from the best fitting quenching value of 360\,ps for 3$^{\mathrm{rd}}$ positive system radiative state for known $E/N$ = 125\,Td in the APTD (as determined using the simplest equivalent circuit \cite{naude2005electrical}) is given as well. It is for $\alpha$ = 0.0043 with ratio value of 0.49 in the APTD, see further in the next section.

\section{\label{experiment} Evaluation of the resulting $E/N$ and discussion}

Using the ratio functions $R_{t-d}$ in Fig.\,\ref{ratioss} and $R_{s-s}$ in \ref{ratiossss}, together with measured optical emission spectra intensities for vibrational bands 
of spectral bands of interest,  
we obtained following values of the ratios and electric fields based on the developed simple models.

\begin{figure}[hbt]
\begin{center}
a)
\includegraphics[width=0.44\textwidth]{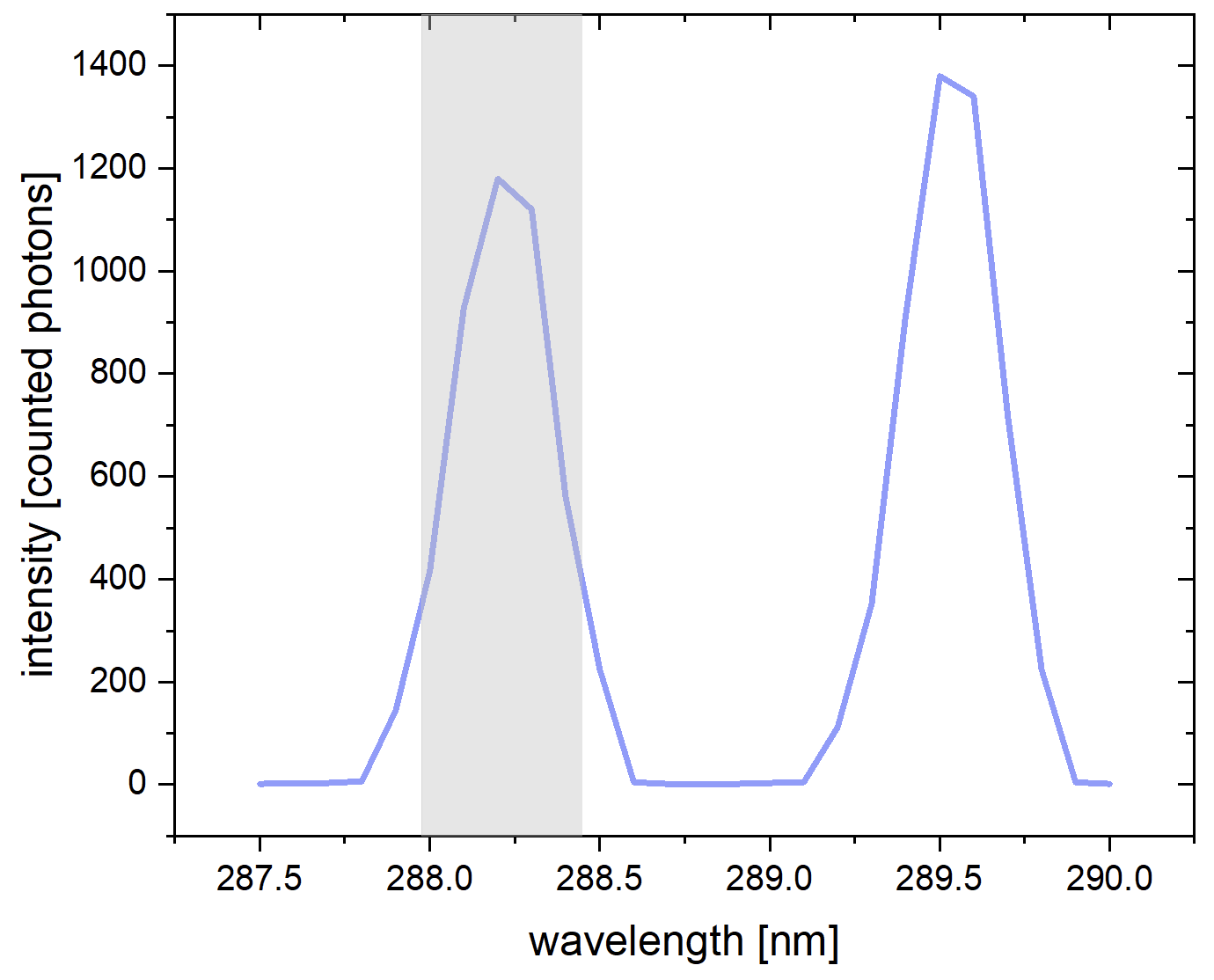}
b)
\includegraphics[width=0.44\textwidth]{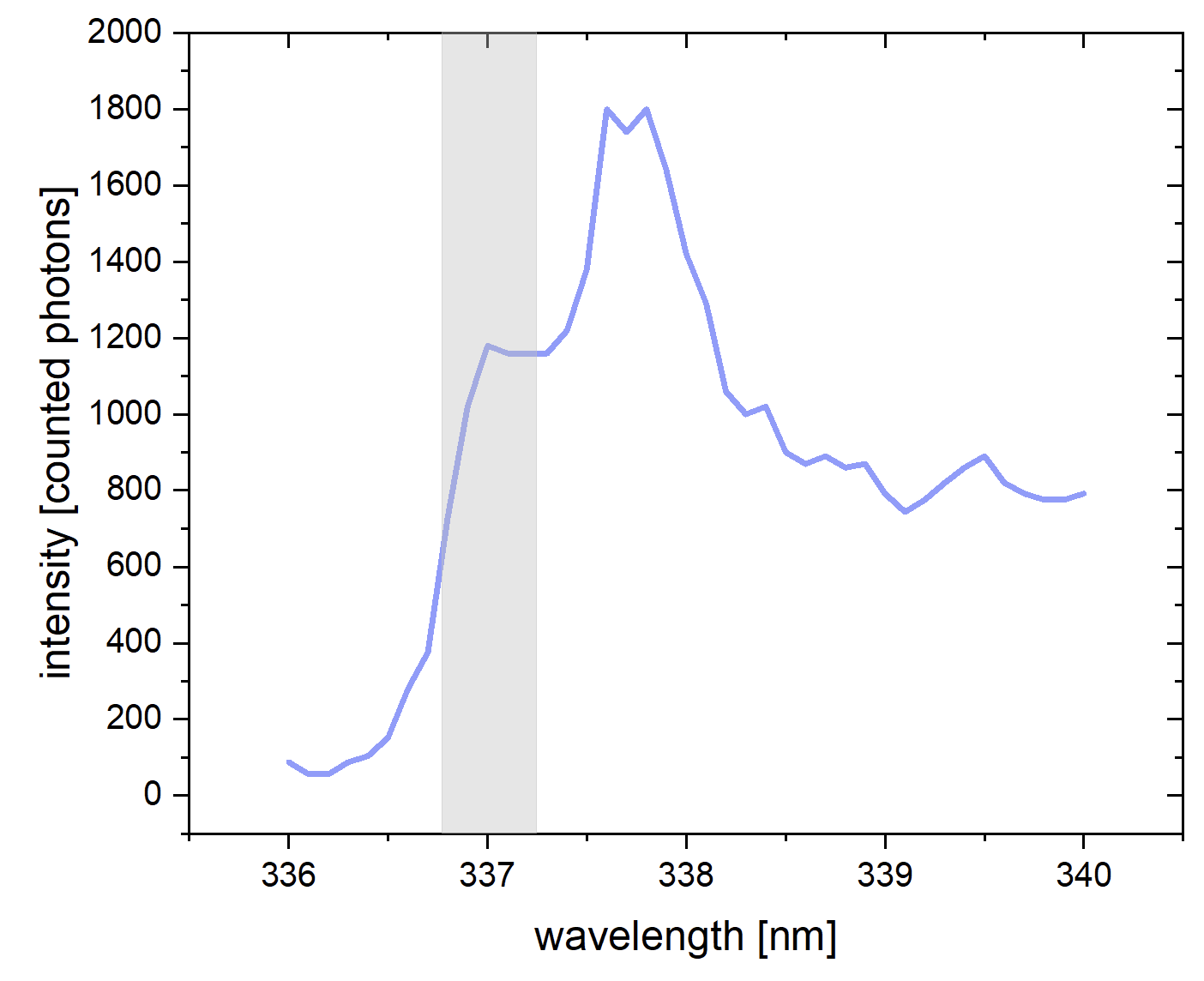}
\caption{The overview spectra and the selection of the spectral intervals for band intensity measurements using TCSPC. In part a), the selection from UV-doublet system is shown using a grey band. In part b), the selection from FDB radiative transition (1$\rightarrow$0)$_{3/2}$ at 337\,nm is shown.}
\label{spectral}
\end{center}
\end{figure}

In the case of the single-filament discharge, the time development of intensity waveforms of the (0$\rightarrow$0) spectral band of UV-doublet and (1$\rightarrow$0)$_{3/2}$ of FDB were recorded using a TCSPC photomultipliers with spectral resolution of approx. 0.46\,nm (measured instrumental function, FWHM). The spectral intervals taken into account for the TCSPC measurements are shown in Fig.\,\ref{spectral}. The intensities were corrected on this spectral selection. As the spectral bands may change in time during the passage of the streamer front or due to the possible overlap with neighboring bands, this spectral selection is source of non-negligible uncertainty which can be corrected by a proper fit of the relevant spectra in future analysis.

\begin{figure}[hbt]
\begin{center}
a)
\includegraphics[width=0.45\textwidth]{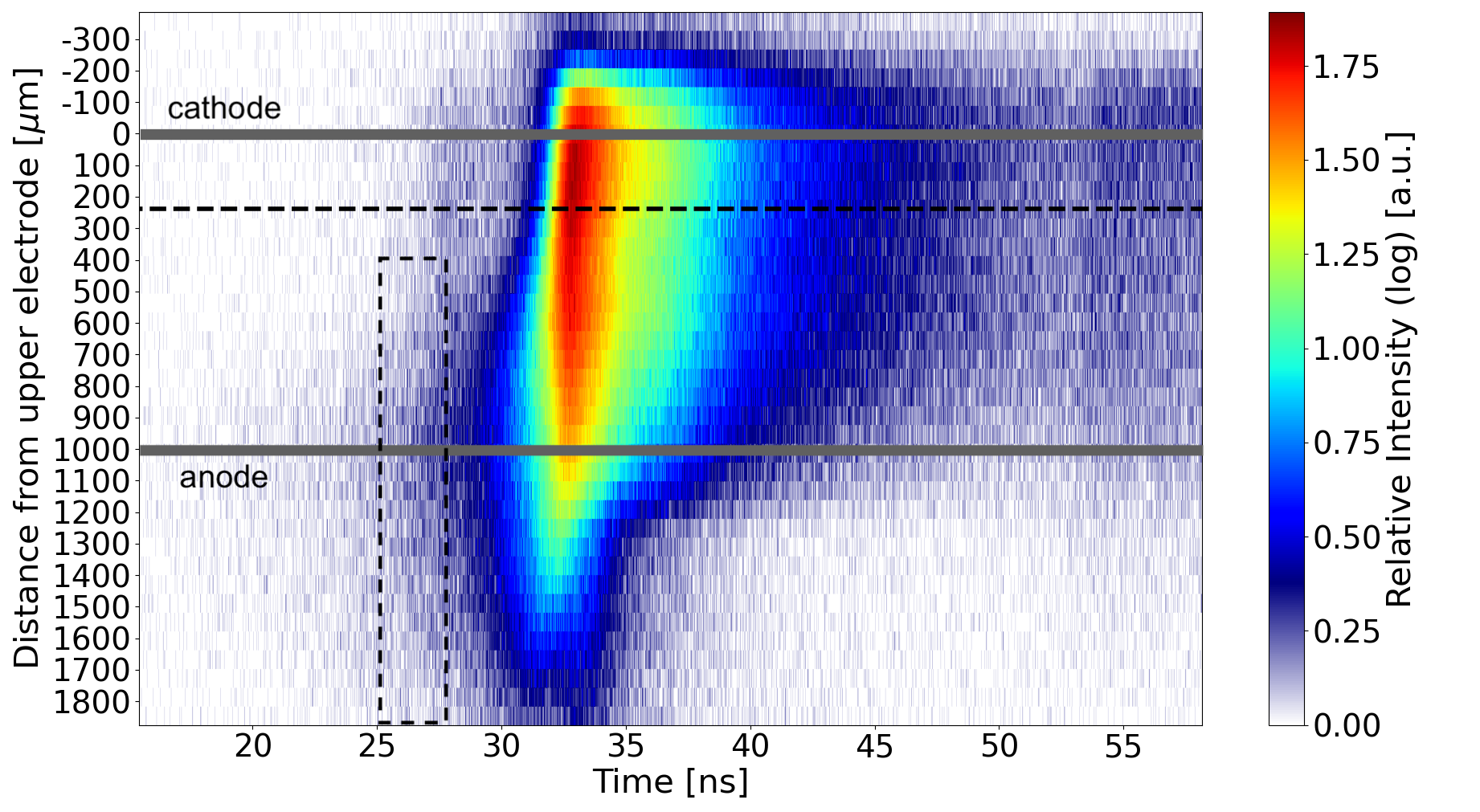}
b)
\includegraphics[width=0.45\textwidth]{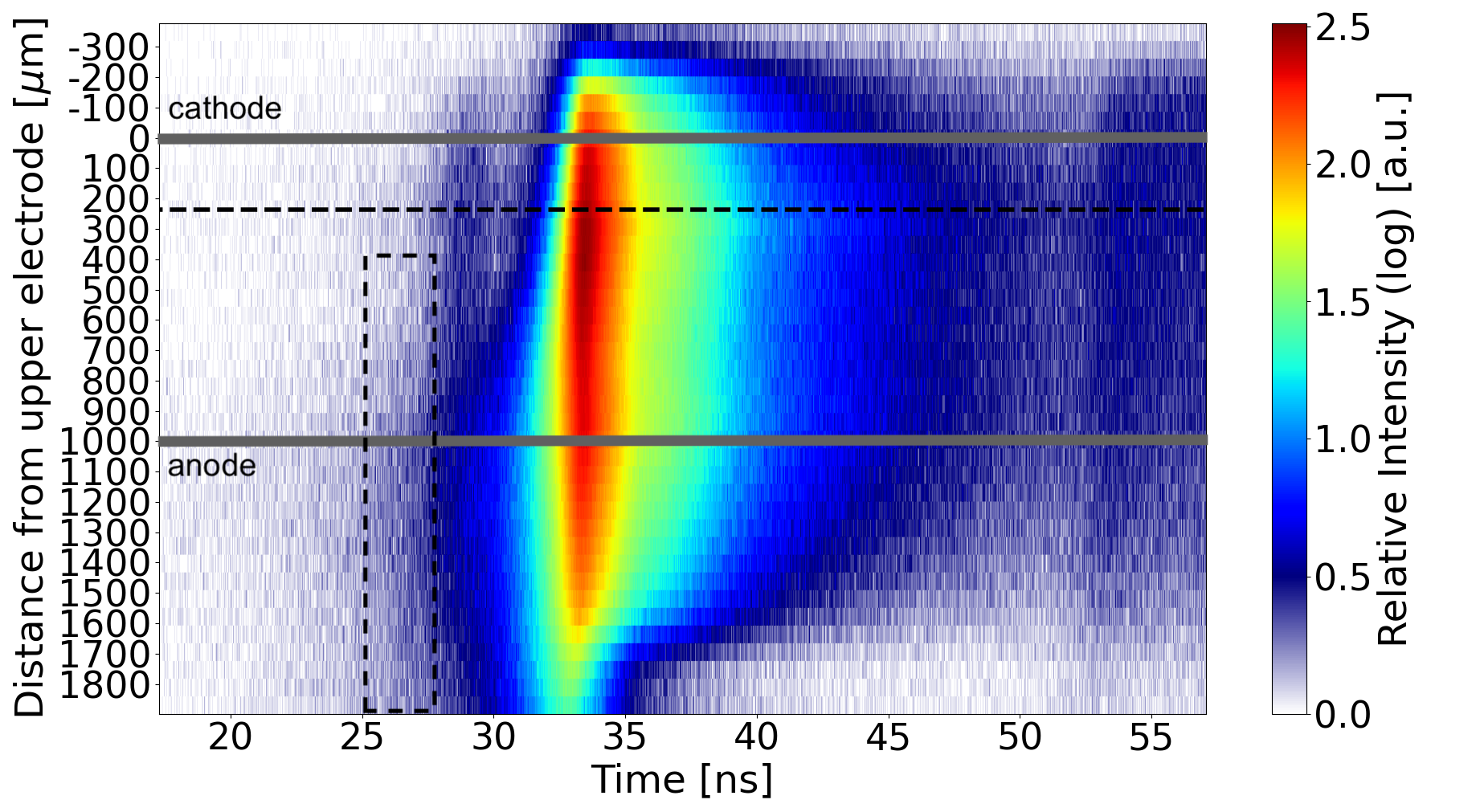}
\caption{The spatiotemporal distribution of the Fox-Duffendack-Baker system emission of the band (1$\rightarrow$0)$_{3/2}$ at 337.0\,nm for single-filament barrier discharge, in part a), and of the UV-doublet (0$\rightarrow$0) emission at 288\,nm, part b). The thick grey lines denote the surface of the dielectrics on corresponding electrodes. The dashed line denotes point of interest for further analysis for intensity ratio method, see Fig.\,\ref{result1}. The dashed rectangle denotes the area used for first Townsend coefficient fitting for $E/N$ determination. The resulting values of the $E/N$ from determined first Townsend coefficients from both systems' spatiotemporal emissions were in agreement.}
\label{scan}
\end{center}
\end{figure}

The results of the TCSPC spatial scan of the inter-electrode distance are shown in Fig.\,\ref{scan}, 5 minutes accumulation for each position.  The spatiotemporal distributions of the FDB emission at 337\,nm and of UV-doublet at 288\,nm are shown there. 
A typical spatiotemporal distribution of the optical emission from single filament dielectric barrier discharge is observed, compare \cite{kozlov2001,jahanbakhsh2019correlation}. 
The intensity of the emission rises from the noise approximately at 25\,ns close to the anode where the electron multiplication takes place and the Townsend pre-phase occurres. Later, around 30\,ns, the streamer starts to propagate towards the cathode and accelerates. It crosses the gap in about 1\,ns, a comparable time with those known from the barrier discharge in air. After the streamer impacts the dielectrics at the cathode, the surface charges deposit onto the dielectrics, decrease the effective electric field in the gap and the discharge is quenched. The intensity of the optical emission diminishes and disappears in the noise at about 50\,ns.

In Fig.\,\ref{scan}a) and b), the dashed line shows coordinate selected for long-accumulation (20 minutes for each wavelength, remeasured two times) measurements of band intensities for further analysis using the ratio function $R_{t-d}$, as shown in Fig.\,\ref{ratioss}b). 
The coordinate is 240\,$\mu$m far away from the cathode. 
It therefore catches the propagating streamer front in close distance from the cathode where its field should be rather high. 
It is approximately at the time of 32\,ns.

\begin{figure}[hbt]
\begin{center}
a)
\includegraphics[width=0.45\textwidth]{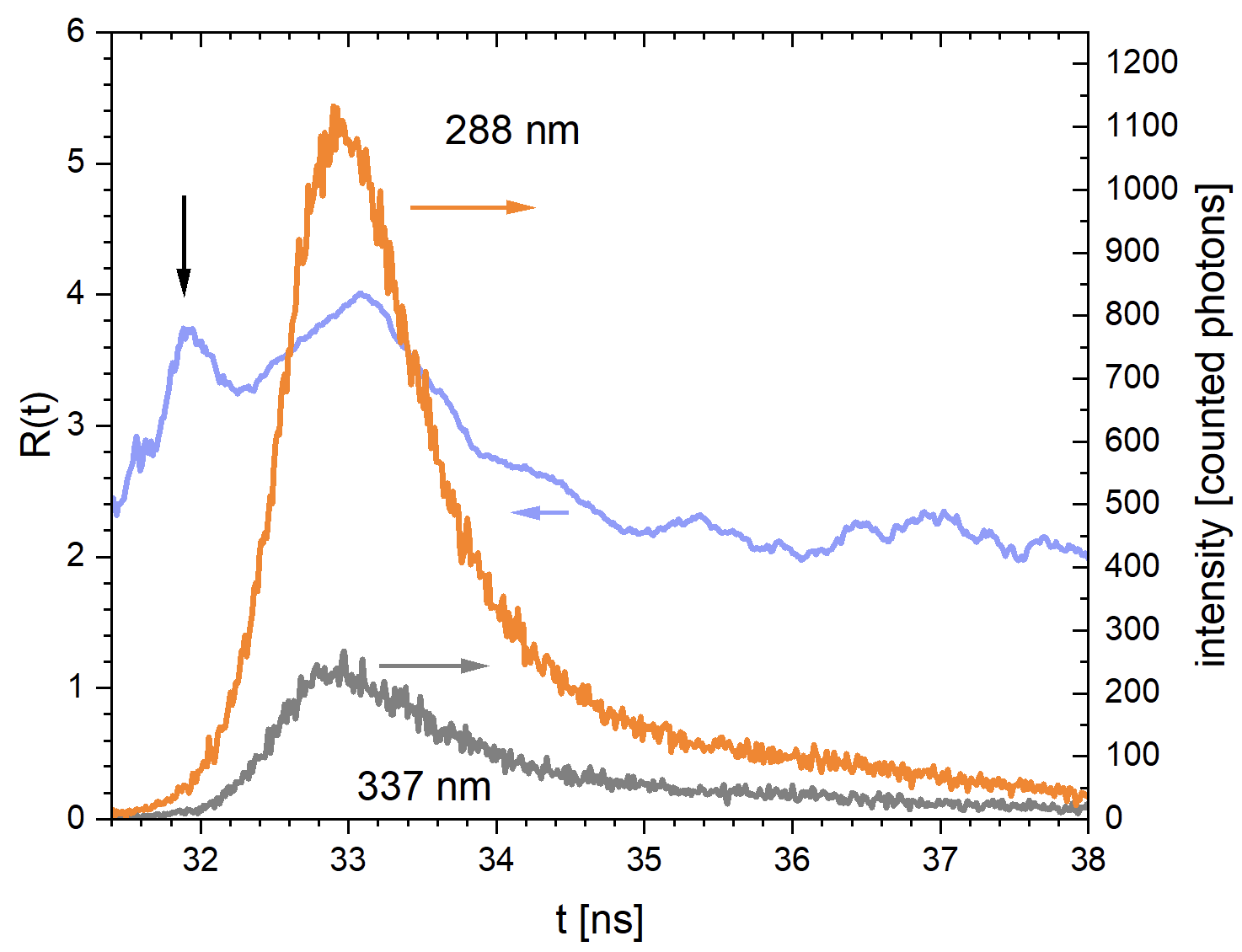}
b)
\includegraphics[width=0.45\textwidth]{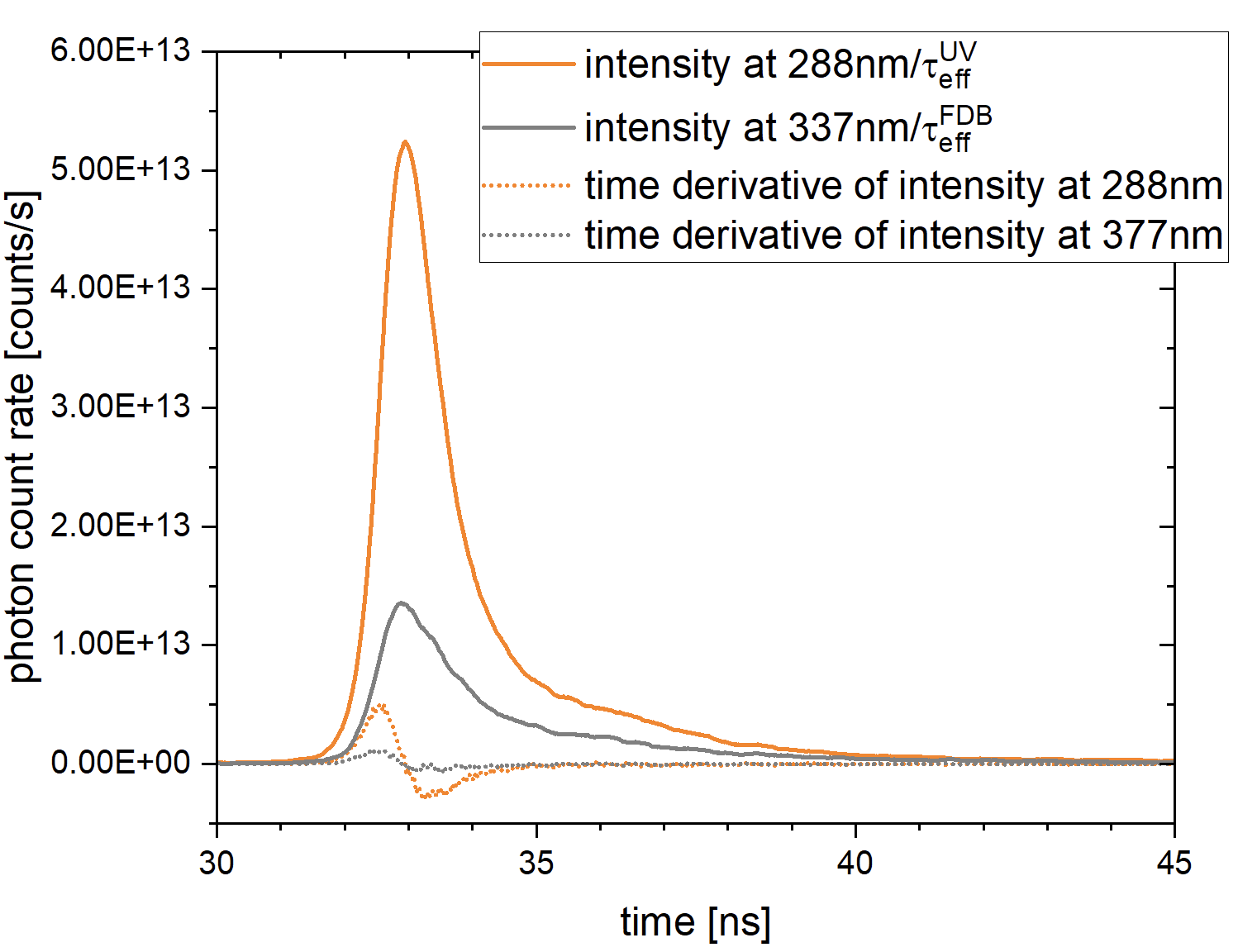}
\caption{The measured intensities of spectral bands UV-doublet (0$\rightarrow$0) at 288\,nm (orange) and FDB (1$\rightarrow$0)$_{3/2}$ at 337\,nm (grey) and the resulting ratio $R(E(t)/N)$ (blue) development for $E/N$ determination, see part a). The peak of the ratio value corresponding to the streamer head with expected highest $E/N$ is denoted with black arrow. In part b), the detail of the constituents of the $R_{t-d}$ function given by the left side of the equation (\ref{master}) is shown for closer analysis. Here, the intensities are smoothed and  corrected on the spectral selection. 
}
\label{result1}
\end{center}
\end{figure}

The result of the long-accumulation measurements is shown in Fig.\,\ref{result1}a). Where besides the intensities of the UV-doublet and FDB bands, the resulting ratio function $R_{t-d}$ is shown too (according to the equation (\ref{master})). The ratio is computed based on data discussed in previous section in tables\,\ref{tableCO2A} and \ref{tableCO2B}. 
The peak value of the ratio at approx. 31.8\,ns is highlighted by an arrow and is at the instant of streamer front passage. The preceding noisy signal is well known and comes from the fact that at that time the intensity values are very small. 
Additionally, the second peak at approx. 33.5\,ns is created due to the step-vise excitation of the higher state (UV-doublet in our case) and therefore not by an electron impact from the ground state as the method requires. It is therefore an artefact.  
In Fig.\,\ref{result1}b), a closer look into the constituents of the $R_{t-d}$ function is given. In this case, the derivatives are non-negligible (about 15\% for UV-doublet and 11\% for FDB at the time of their peak values), yet not larger than the intensities divided by the effective lifetimes. It is similar situation as in the case of the first negative system signal in air which has also a high energy threshold of the radiative state, see \cite{goldberg2022}.

At this point of the analysis, it is clear that our measurements results in a typical ratio waveform as it is known from its application in nitrogen containing plasmas, compare Fig.\,3 in \cite{goldberg2022}. 
It means that the method is sensitive to the electric field peak which can be clearly marked. Note the very small energetic difference between the excitation thresholds of the used radiative states of UV-doublet and FDB. The TCSPC technique is apparently sensitive enough to resolve such a fine changes whether originating in time or intensity. 

The peak value of the ratio function in the denoted  streamer front is approx. 3.7 which, if read out of the curve in Fig.\,\ref{ratioss}, result in the reduced electric field value of approx. 45\,Td. Such a value is un-physically low for a streamer under given conditions. 
Taking into account the results of a fluid models in the literature \cite{ponduri2016}, where the value of the electric field is around 400\,Td. 
Nevertheless, following facts needs to be taken into account. 
The ratio function $R_{t-d}$ presented in Fig.\,\ref{ratioss}b) is of lower sensitivity to the $E/N$ changes as in the case of nitrogen. In this case spanning in interval of one order of magnitude (say from 3 to 14). The sensitivity of the method for a nitrogen plasmas spans over four orders of magnitude \cite{goldberg2018}. 
In future, both an uncertainty quantification  together with an independent $E/N$ measurement for comparison need to be carried out in order to find the proper set of kinetic data, see e.g. \cite{bilek2018electric,mrkvickova2023}.  

\begin{figure}[hbt]
\begin{center}
\includegraphics[width=0.5\textwidth]{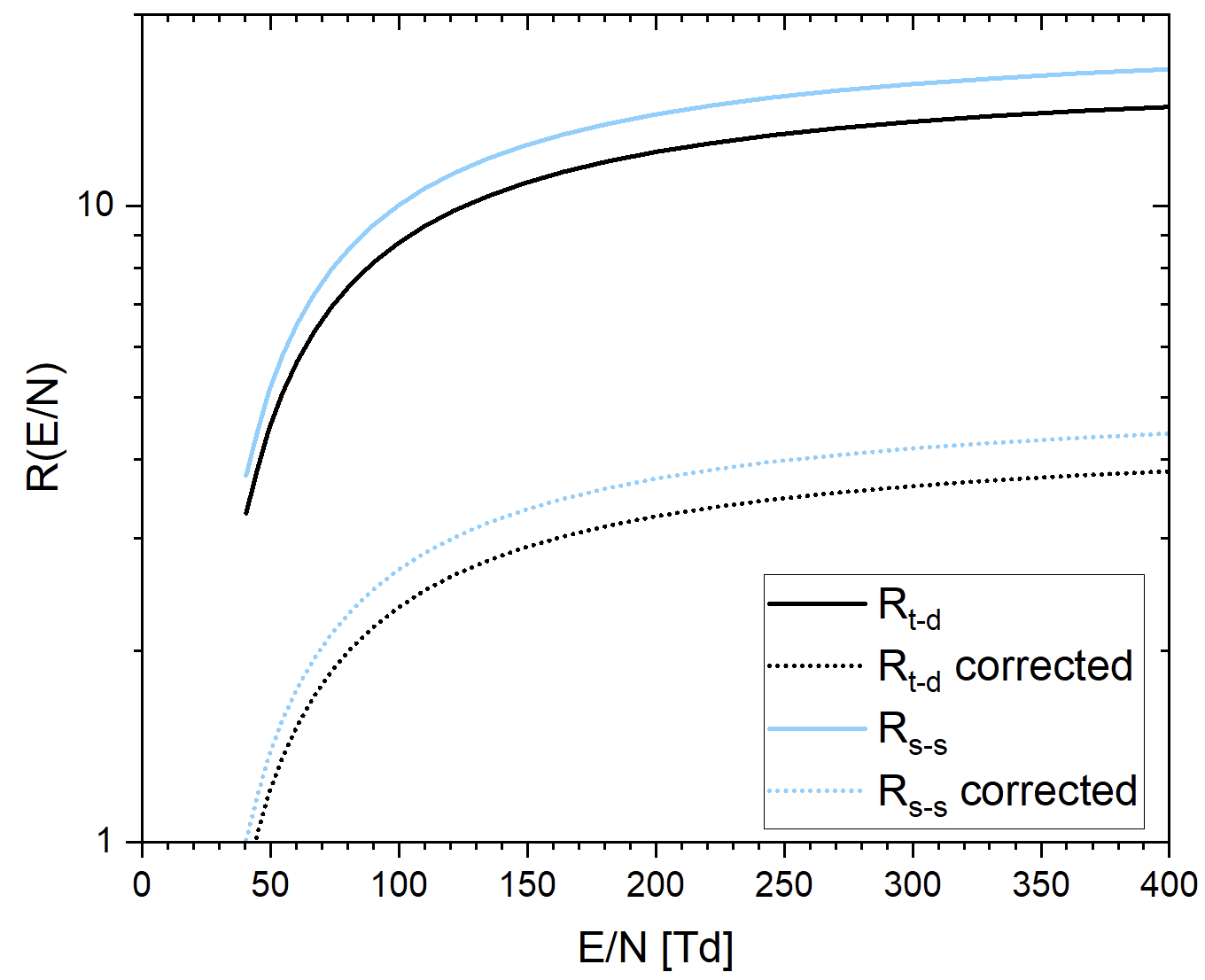}
\caption{The functions $R_{t-d}(E/N)$ (equation\,\ref{master}) and $R_{s-s}(E/N)$ (equation \ref{mastersteady}) for the ratio of UV-doublet (0$\rightarrow$0) and Fox-Duffendack-Baker (1$\rightarrow$0)$_{3/2}$ system bands intensities are presented for sets of excitation Franck-Condon factor values from based on Farley et al., see Fig.\,\ref{ratioss}. The corrected ratio functions are also presented after the calibration on the Townsend discharge pre-phase of the discharge development.}
\label{ccs}
\end{center}
\end{figure}

However, we have done an independent calibration of the ratio function similarly to \cite{kozlov2001, jahanbakhsh2019correlation}. The optical emission close to the anode prior to the streamer generation, where the discharge mechanism is a Townsend electron multiplication (see Fig.\,\ref{scan}), was integrated over time and from the exponential increase towards the anode the value of the first Townsend coefficient for CO$_2$ was determined. 
As the coefficient is an unambiguous function of the electric field, using LXCat computations (Bolsig+ with Biagi CO$_2$ database at 300\,K) a value of approximately 120\,Td was obtained. Comparing to the optical emission intensity ratio in that area, the original ratio function R$_{t-d}$ was then corrected, i.e. divided by factor of 3.74, to fit the $E/N$ value. 
The original as well as the corrected ratio functions for both steady state and time-dependent ratio functions are shown in Fig.\,\ref{ccs}. 
For the corrected time-dependent ratio function, the ratio amplitude of 3.7 gives a very reasonable value of 330\,Td. 
Nevertheless, the cause responsible for the multiplication factor needs to be clarified. 
The most probable candidates are the overlapping spectra of the FDB system, (1$\rightarrow$0)$_{3/2}$ at 337\,nm with neighboring (1$\rightarrow$0)$_{1/2}$ at 338\,nm, imperfection in the optical setup and the kinetic and atomic data uncertainties which are left for future work. 
Finally after the correction, the working intensity ratio method for streamer discharges in CO$_2$ is here presented. 
Similar approach was used in the pioneering work of Kozlov et al. \cite{kozlov2001} for air. 

\begin{figure}[hbt]
\begin{center}
a)
\includegraphics[width=0.45\textwidth]{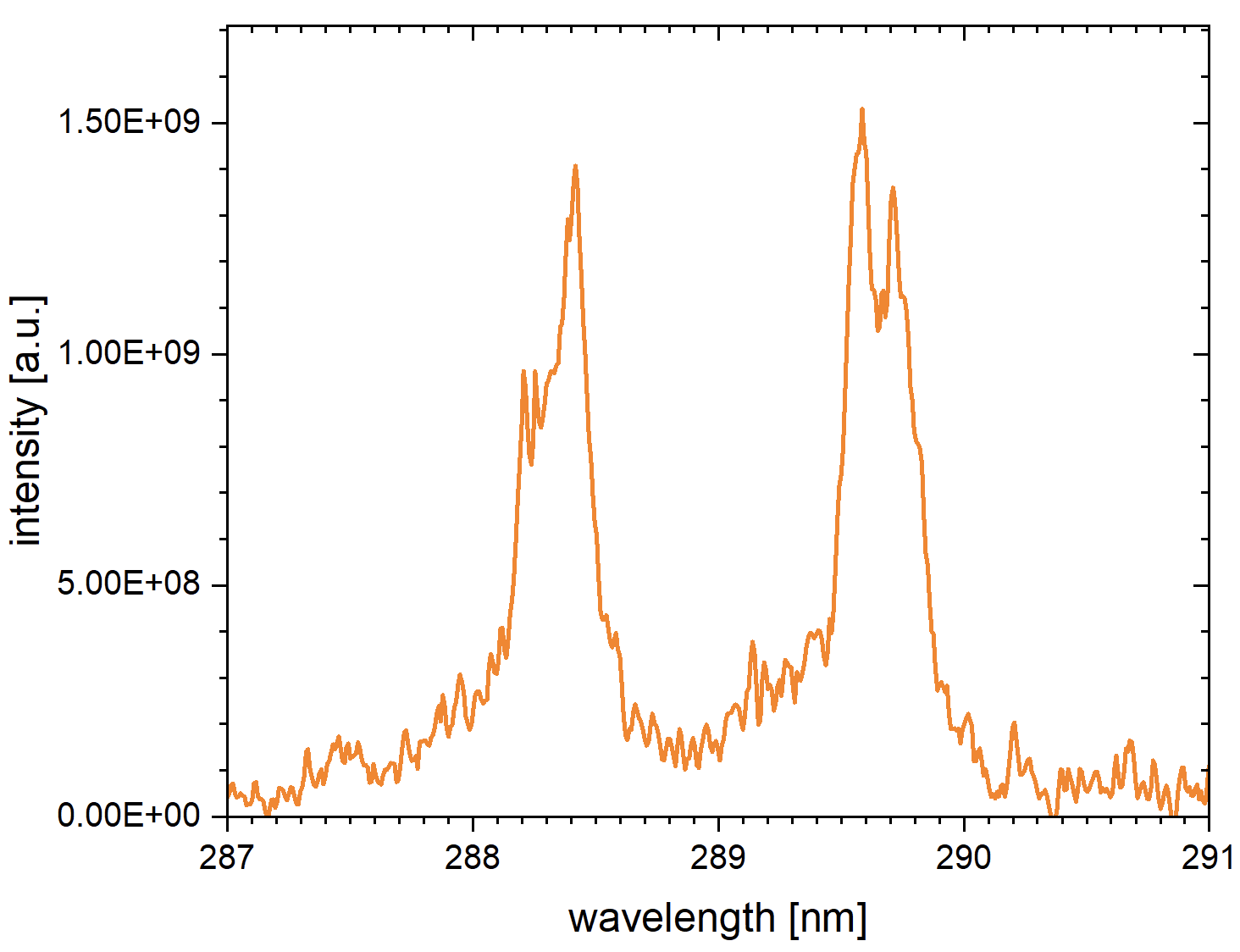}
b)
\includegraphics[width=0.45\textwidth]{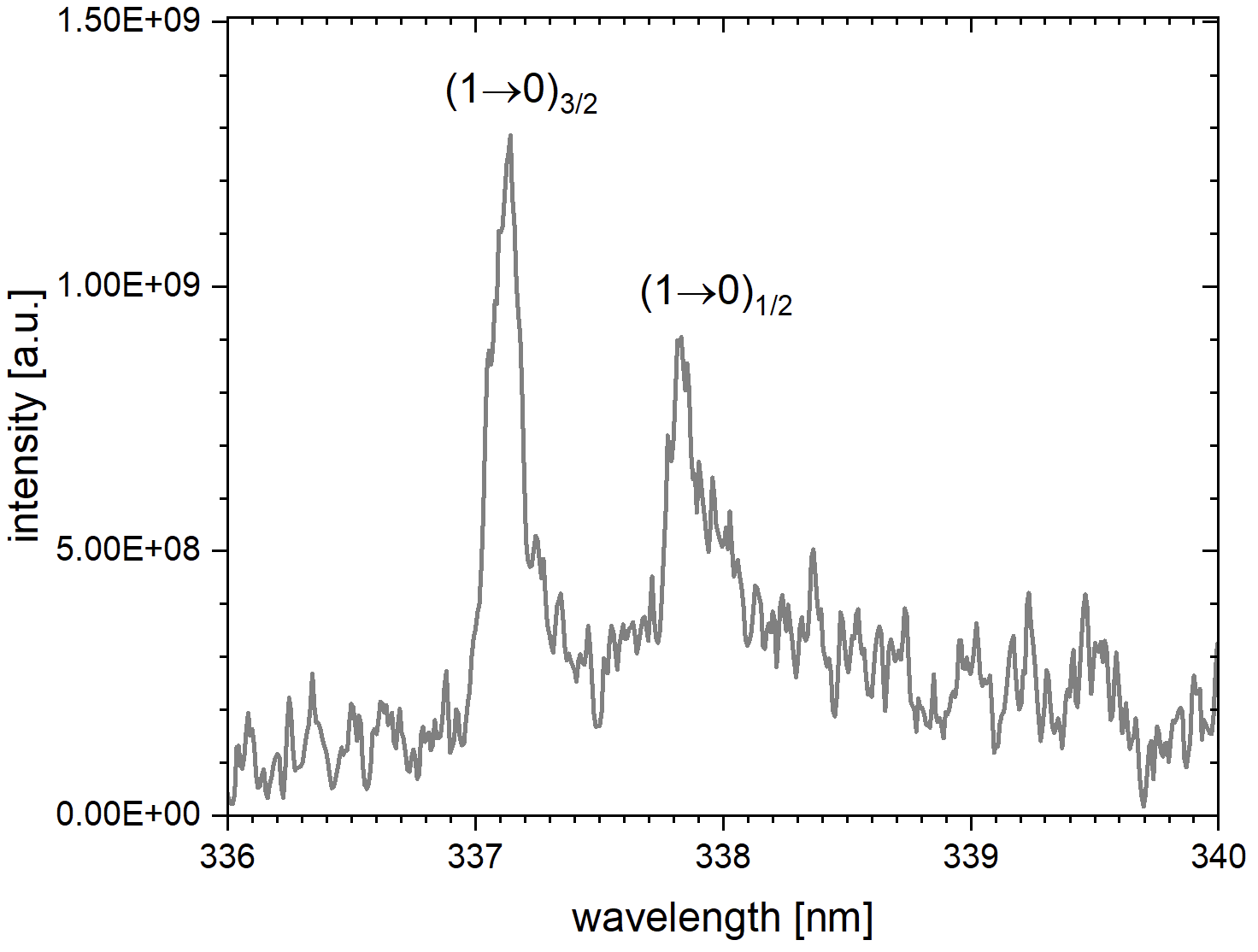}
\caption{The optical emission spectra of UV-doublet (0$\rightarrow$0), part a), and Fox-Duffendack-Baker (1$\rightarrow$0)$_{3/2}$, part b), system bands intensities for APTD at 2\,kHz and 12\,kV$_{pp}$. 
}
\label{townsendSS}
\end{center}
\end{figure}

The optical emission spectra of UV-doublet and FDB were measured using gated ICCD also for APTD, multi-filamentary and single-filament barrier discharge with the same optical setup (ICCD, monochromator and optics) for determination of the time-independent ratios. The spectra for APTD are shown in Fig.\,\ref{townsendSS}. 
The resulting ratio of UV-doublet (0$\rightarrow$0) and Fox-Duffendack-Baker (1$\rightarrow$0)$_{3/2}$ have a value of approximately 2.8 ($\pm$0.1) which using the corrected $R_{s-s}(E/N)$ function from Fig.\,\ref{ccs} results in the $E/N$ of approximately 110\,Td ($\pm$10\,Td). This is a reasonable value, the $E/N$ determined for the discharge by the simplest equivalent circuit is about 125\,Td. Note that for the case of microseconds long homogeneous APTD discharge the steady-state model is an appropriate description and spatially averaged emission measurements are sufficient.

For the multi-filament and single-filament barrier discharge regimes the steady-state model gives only an effective (spatiotemporally averaged) information about the $E/N$ and can't be used for proper diagnostics. Note the nonnegligible contribution of the derivatives to the ratio function mentioned earlier. 
The resulting ratios from similar measurements as shown in Fig.\,\ref{townsendSS} have values of 3.9 for the multi-filamentary discharge (1\,mm gap, 8\,kHz, 18\,kV$_{pp}$) and 5.0 for the single-filament discharge (1\,mm gap, 2\,kHz, 12\,kV$_{pp}$). 
The ratio value of 3.9 results for corrected $R_{s-s}(E/N)$ function in 240\,Td, i.e. underestimates the peak electric field by 30\%. 
The value of 5.0 is over the scale for corrected $R_{s-s}(E/N)$ function. Apparently, the application of the steady state approach has no practical use in these cases.

The last result concerns the use of combined CO (3$^{rd}$ positive system, (0$\rightarrow$0) transition at 283\,nm) and CO$_2^+$ (UV-doublet system, (0$\rightarrow$0) transition at 288\,nm) spectral bands for diagnostics of the APTD discharge. The spectra were introduced in Fig.\,\ref{spectra-alpha}.  
For this analysis, the ratio functions $R_{s-s}$ are shown in Fig.\,\ref{ratiossss}. 
The spectra depicted in Fig.\,\ref{spectra-alpha} give following ratios: 0.49 for 7.4\,sccm with $\alpha$ = 0.0043, 1.77 for 37\,sccm with $\alpha$ = 0.00183 and 1.89 for 74\,sccm with $\alpha$ = 0.00117. 
Such ratio values results using the full line $R_{s-s}$ functions from Fig.\,\ref{ratiossss} in $E/N$ values of 104, 111 and 104\,Td, respectively. 
Such values are lower with respect to the $E/N$ determined for given discharges using equivalent electrical circuit \cite{naude2005electrical}, 125\,Td for all flow rates. The equivalent circuit showed to be relatively precise method for APTD $E/N$ determination, see \cite{mrkvickova2023}.

Even though the whole collisional-radiative model relies on multiple fundamental data with their own uncertainties, we have done additional estimation to quantify the quenching rate coefficient of the $\mathrm{CO(b}^3\Sigma)(\nu' = 0)$ state by CO$_2$.  
Modifying the $\tau_{eff}^{3rd}$ value, until the ratio function $R_{s-s}$ (see the dashed line in Fig.\,\ref{ratiossss}) leads to $E/N$ value which is in agreement with the result of equivalent electrical circuit, i.e. 125\,Td, we obtain effective lifetime of $\tau_{eff}^{3rd}$ = 360\,ps. 
Taking into account the radiative lifetime of 57.6\,ns and the equation (\ref{effective}) we estimate the rate coefficient to be of 1.15$\cdot10^{-10}$ cm$^3$/s (assuming temperature 300\,K, i.e. n$_{CO_2}$=2.414$\cdot10^{19}$\,cm$^{-3}$, the admixture of CO has negligible influence within this estimate). 
Finally, the combination of these two bands results in very sensitive ratio function (spanning over six orders of magnitude for low $E/N$ values) and is certainly suitable for further improvement.

\section{\label{conclusion} Summary and conclusion}

This article presents a detailed optical emission spectroscopic study of three barrier discharges operated in atmospheric pressure CO$_2$: APTD, multi-filamentary and single-filament barrier discharge. Applying several monochromators and detectors, we were able to record the spectra for a wide spectral interval from 250 to 850\,nm and resolve and identify the main spectral systems and atomic lines occurring in these plasmas. 

We made an attempt to find suitable spectral signatures for determination of the $E/N$ in the CO$_2$ plasmas, i.e. in CO/CO$_2$ mixtures, using the intensity ratio method. 
For the first time with such aim, several spectral systems were investigated using available data in literature, Bolsig+ solver and the LXCat database and some collision-radiative models were proposed. 
Even though some of the resulting $E/N$ values differ from those known from literature or measured by other means, the values are in reasonable range. 
The radiative states responsible for spectral bands occurring in the spectra of investigated discharge are not ideal for such method, their excitation energy threshold differences are rather small, as in contrast to air plasmas and the use of second positive and first negative systems of nitrogen molecule. 
For single-filament streamer barrier discharge, the TCSPC technique proves its excellent properties regarding its high temporal resolution, sensitivity as well as large dynamic range to show applicability for the streamer head detection in time. 

Multiple uncertainties originating from different electron-molecule interaction cross-sections, quenching rates, excitation Franck-Condon factors or radiative lifetimes were identified as possible causes for the overall method uncertainty. Some missing values were reported as well which need to be added in the future. We have also  estimated some unknown values for effective lifetimes of some radiative states. 

The intensity ratio methods were prepared both for time-dependent as well as steady-state case of discharges under investigation. 
The time-dependent model for the $E/N$ determination in streamer barrier discharge using TCSPC technique and intensity ratio of UV doublet (0$\rightarrow$0) and FDB (1$\rightarrow$0)$_{3/2}$ bands resulted in reasonable values - for kHz repetitive barrier discharge. 
Here developed UVdoublet/FDB time-dependent ratio method is challenging yet applicable result of our analysis for $E/N$ determination in streamer CO$_2$ discharges, similarly as it was done for air earlier. 
The same corrected ratio was used also for the APTD with good quantitative result. 
Nevertheless, the  uncertainty and its sources still need to be quantified and identified for easier and reliable utilization. 

Moreover, the approach for $E/N$ determination using spectral bands coming from radiative states originating from electron excitation of two different ground state molecules is utilized and discussed for the first time. It is based on the emission of the 3$^{rd}$ positive system of CO and UV-doublet system of the CO$_2$ molecule. It is applied to the APTD discharge yet slightly underestimating the resulting $E/N$ values. To find the cause of such a difference can be important for CO$_2$ plasma kinetics in general, especially to clarify the values of the effective lifetimes under given conditions.

Finally, despite some challenging issues in the task, the first important steps in development of intensity ratio method for $E/N$ determination in CO$_2$ plasmas were made here and we have clearly identified the imperfections and show the necessary next steps to be made.

\section*{Acknowledgements}

The authors are grateful to Luk\'a\v s Kus\'yn for his technical contribution. This research was supported by the project Nr. TK04020069 funded by the Technology Agency of the Czech Republic as a part of the Programme TH\'ETA, MOBILITY programme 8J23FR008 funded by the Ministry of Education, Youth and Sports of the Czech Republic, 
by the Barrande mobility programm of Campus France funded by the French Ministry of Europe and foreign affairs and by the French Ministry of higher education and research, 
and by project CEPLANT (LM2023039) funded by the Ministry of Education, Youth and Sports of the Czech Republic.

\section*{References}
\bibliography{references}

\providecommand{\newblock}{}
\begin{thebibliography}{10}
\expandafter\ifx\csname url\endcsname\relax
  \def\url#1{{\tt #1}}\fi
\expandafter\ifx\csname urlprefix\endcsname\relax\def\urlprefix{URL }\fi
\providecommand{\eprint}[2][]{\url{#2}}

\bibitem{kogelschatz2003}
Kogelschatz U 2003 {\em Plasma Chemistry and Plasma Processing\/} {\bf 23}
  1--46 \urlprefix\url{https://doi.org/10.1023/A:1022470901385}

\bibitem{brandenburg2017}
Brandenburg R 2017 {\em Plasma Sources Science and Technology\/} {\bf 26}
  053001 \urlprefix\url{https://doi.org/10.1088/1361-6595/aa6426}

\bibitem{eliasson1987}
Eliasson B, Hirth M and Kogelschatz U 1987 {\em Journal of Physics D: Applied
  Physics\/} {\bf 20} 1421--1437
  \urlprefix\url{https://doi.org/10.1088/0022-3727/20/11/010}

\bibitem{kozlov2000}
Kozlov K~V, Michel P and Wagner H~E 2000 {\em Plasmas and Polymers\/} {\bf 5}
  129--150 \urlprefix\url{https://doi.org/10.1023/A:1011302017354}

\bibitem{kraus2002}
Kraus M, Egli W, Haffner K, Eliasson B, Kogelschatz U and Wokaun A 2002 {\em
  Phys. Chem. Chem. Phys.\/} {\bf 4}(4) 668--675
  \urlprefix\url{http://dx.doi.org/10.1039/B108040G}

\bibitem{corke2010}
Corke T~C, Enloe C~L and Wilkinson S~P 2010 {\em Annual Review of Fluid
  Mechanics\/} {\bf 42} 505--529 (\textit{Preprint}
  \eprint{https://doi.org/10.1146/annurev-fluid-121108-145550})
  \urlprefix\url{https://doi.org/10.1146/annurev-fluid-121108-145550}

\bibitem{cernak2011}
Černák M, Kováčik D, Ráhel' J, St'ahel P, Zahoranová A, Kubincová J,
  Tóth A and Černáková L 2011 {\em Plasma Physics and Controlled Fusion\/}
  {\bf 53} 124031
  \urlprefix\url{https://dx.doi.org/10.1088/0741-3335/53/12/124031}

\bibitem{brandt2016}
Brandt S, Schütz A, Klute F, Kratzer J and Franzke J 2016 {\em Spectrochimica
  Acta Part B: Atomic Spectroscopy\/} {\bf 123} 6--32 ISSN 0584-8547
  \urlprefix\url{https://www.sciencedirect.com/science/article/pii/S0584854716300969}

\bibitem{pietanza2021}
Pietanza L~D, Guaitella O, Aquilanti V, Armenise I, Bogaerts A, Capitelli M,
  Colonna G, Guerra V, Engeln R, Kustova E, Lombardi A, Palazzetti F and Silva
  T 2021 {\em The European Physical Journal D\/} {\bf 75} 237
  \urlprefix\url{https://doi.org/10.1140/epjd/s10053-021-00226-0}

\bibitem{brehmer2014}
Brehmer F, Welzel S, van~de Sanden M~C~M and Engeln R 2014 {\em Journal of
  Applied Physics\/} {\bf 116} 123303 ISSN 0021-8979 (\textit{Preprint}
  \eprint{https://pubs.aip.org/aip/jap/article-pdf/doi/10.1063/1.4896132/15145184/123303\_1\_online.pdf})
  \urlprefix\url{https://doi.org/10.1063/1.4896132}

\bibitem{belov2016}
Belov I, Paulussen S and Bogaerts A 2016 {\em Plasma Sources Science and
  Technology\/} {\bf 25} 015023
  \urlprefix\url{https://dx.doi.org/10.1088/0963-0252/25/1/015023}

\bibitem{ponduri2016}
Ponduri S, Becker M~M, Welzel S, van~de Sanden M~C~M, Loffhagen D and Engeln R
  2016 {\em Journal of Applied Physics\/} {\bf 119} 093301 ISSN 0021-8979
  (\textit{Preprint}
  \eprint{https://pubs.aip.org/aip/jap/article-pdf/doi/10.1063/1.4941530/13435447/093301\_1\_online.pdf})
  \urlprefix\url{https://doi.org/10.1063/1.4941530}

\bibitem{brandenburg2017b}
Brandenburg R and Sarani A 2017 {\em The European Physical Journal Special
  Topics\/} {\bf 226} 2911--2922
  \urlprefix\url{https://doi.org/10.1140/epjst/e2016-60339-8}

\bibitem{douat2023}
Douat C, Ponduri S, Boumans T, Guaitella O, Welzel S, Carbone E and Engeln R
  2023 {\em Plasma Sources Science and Technology\/} {\bf 32} 055001
  \urlprefix\url{https://dx.doi.org/10.1088/1361-6595/acceca}

\bibitem{bajon2023}
Bajon C, Dap S, Belinger A, Guaitella O, Hoder T and Naudé N 2023 {\em Plasma
  Sources Science and Technology\/} {\bf 32} 045012
  \urlprefix\url{https://dx.doi.org/10.1088/1361-6595/acc9d9}

\bibitem{snoeckx2017}
Snoeckx R and Bogaerts A 2017 {\em Chem. Soc. Rev.\/} {\bf 46}(19) 5805--5863
  \urlprefix\url{http://dx.doi.org/10.1039/C6CS00066E}

\bibitem{seeger2015}
Seeger M 2015 {\em Plasma Chemistry and Plasma Processing\/} {\bf 35} 527--541
  \urlprefix\url{https://doi.org/10.1007/s11090-014-9595-4}

\bibitem{seeger2017}
Seeger M, Avaheden J, Pancheshnyi S and Votteler T 2016 {\em Journal of Physics
  D: Applied Physics\/} {\bf 50} 015207
  \urlprefix\url{https://dx.doi.org/10.1088/1361-6463/50/1/015207}

\bibitem{rabie2018}
Rabie M and Franck C~M 2018 {\em Environmental Science \& Technology\/} {\bf
  52} 369--380 \urlprefix\url{https://doi.org/10.1021/acs.est.7b03465}

\bibitem{vass2021}
Vass M, Egüz E, Chachereau A, Hartmann P, Korolov I, Hösl A, Bošnjaković D,
  Dujko S, Donkó Z and Franck C~M 2020 {\em Journal of Physics D: Applied
  Physics\/} {\bf 54} 035202
  \urlprefix\url{https://dx.doi.org/10.1088/1361-6463/abbb07}

\bibitem{rankovic2020}
Ranković M, Kumar T~P R, Nag P, Kočišek J and Fedor J 2020 {\em The Journal
  of Chemical Physics\/} {\bf 152} 244304 ISSN 0021-9606 (\textit{Preprint}
  \eprint{https://pubs.aip.org/aip/jcp/article-pdf/doi/10.1063/5.0008897/13897246/244304\_1\_online.pdf})
  \urlprefix\url{https://doi.org/10.1063/5.0008897}

\bibitem{vemu2023}
Vemulapalli H and Franck C~M 2023 {\em Journal of Physics D: Applied Physics\/}
  {\bf 56} 065202 \urlprefix\url{https://dx.doi.org/10.1088/1361-6463/acaab7}

\bibitem{goldberg2022}
Goldberg B~M, Hoder T and Brandenburg R 2022 {\em Plasma Sources Science and
  Technology\/}
  \urlprefix\url{http://iopscience.iop.org/article/10.1088/1361-6595/ac6e03}

\bibitem{kozlov2001}
Kozlov K~V, Wagner H~E, Brandenburg R and Michel P 2001 {\em Journal of Physics
  D: Applied Physics\/} {\bf 34} 3164--3176
  \urlprefix\url{https://doi.org/10.1088/0022-3727/34/21/309}

\bibitem{gharib2017}
Gharib M, Mendoza S, Rosenfeld M, Beizai M and Pereira F~J~A 2017 {\em
  Proceedings of the National Academy of Sciences\/} {\bf 114} 12657--12662
  (\textit{Preprint}
  \eprint{https://www.pnas.org/doi/pdf/10.1073/pnas.1712717114})
  \urlprefix\url{https://www.pnas.org/doi/abs/10.1073/pnas.1712717114}

\bibitem{bilek2019}
Bílek P, Šimek M and Bonaventura Z 2019 {\em Plasma Sources Science and
  Technology\/} {\bf 28} 115011
  \urlprefix\url{https://dx.doi.org/10.1088/1361-6595/ab3936}

\bibitem{jansky2021}
J{\'a}nsk{\'y} J, Bessi{\'e}res D, Brandenburg R, Paillol J and Hoder T 2021
  {\em Plasma Sources Science and Technology\/} {\bf 30} 105008

\bibitem{dijcks2022}
Dijcks S, Kusyn L, Janssen J, Bilek P, Nijdam S and Hoder T 2023 {\em Frontiers
  in Physics\/} {\bf 11} ISSN 2296-424X
  \urlprefix\url{https://www.frontiersin.org/articles/10.3389/fphy.2023.1120284}

\bibitem{goekce2016}
Goekce S, Peschke P, Hollenstein C, Leyland P and Ott P 2016 {\em Plasma
  Sources Science and Technology\/} {\bf 25} 045002
  \urlprefix\url{https://dx.doi.org/10.1088/0963-0252/25/4/045002}

\bibitem{siepa2014}
Siepa S, Danko S, Tsankov T~V, Mussenbrock T and Czarnetzki U 2014 {\em Journal
  of Physics D: Applied Physics\/} {\bf 47} 445201
  \urlprefix\url{https://dx.doi.org/10.1088/0022-3727/47/44/445201}

\bibitem{dyatko2021}
Dyatko N~A, Ionikh Y~Z and Meshchanov A~V 2021 {\em Plasma Sources Science and
  Technology\/} {\bf 30} 055015
  \urlprefix\url{https://dx.doi.org/10.1088/1361-6595/abda9e}

\bibitem{kusyn2023}
Kusýn L, Prokop D, Navrátil Z and Hoder T 2023 {\em Plasma Sources Science
  and Technology\/} {\bf 32} 045006
  \urlprefix\url{https://dx.doi.org/10.1088/1361-6595/acc6eb}

\bibitem{rosny1983}
de~Rosny G, Mosburg Earl~R J, Abelson J~R, Devaud G and Kerns R~C 1983 {\em
  Journal of Applied Physics\/} {\bf 54} 2272--2275 ISSN 0021-8979
  (\textit{Preprint}
  \eprint{https://pubs.aip.org/aip/jap/article-pdf/54/5/2272/7981611/2272\_1\_online.pdf})
  \urlprefix\url{https://doi.org/10.1063/1.332381}

\bibitem{kokubo1990}
Kokubo T, Tochikubo F and Makabe T 1990 {\em Applied Physics Letters\/} {\bf
  56} 818--820 ISSN 0003-6951 (\textit{Preprint}
  \eprint{https://pubs.aip.org/aip/apl/article-pdf/56/9/818/7776432/818\_1\_online.pdf})
  \urlprefix\url{https://doi.org/10.1063/1.103320}

\bibitem{ushita1967}
Ushita T, Ikuta N and Yatsuzuka M 1967 Fast time analysis of negative pulse
  corona in air {\em Bulletin of Faculty of Engineering Tokushima University
  (Tokushima, Japan)\/} vol~4 p~89

\bibitem{gravendeel1988}
Gravendeel B, de~Hoog F~J and Schoenmakers M~A~M 1988 {\em Journal of Physics
  D: Applied Physics\/} {\bf 21} 744
  \urlprefix\url{https://dx.doi.org/10.1088/0022-3727/21/5/012}

\bibitem{kondo1980}
Kondo K and Ikuta N 1980 {\em Journal of Physics D: Applied Physics\/} {\bf 13}
  L33 \urlprefix\url{https://dx.doi.org/10.1088/0022-3727/13/2/003}

\bibitem{shcherbakov2007}
Shcherbakov Y~V and Sigmond R~S 2007 {\em Journal of Physics D: Applied
  Physics\/} {\bf 40} 460
  \urlprefix\url{https://dx.doi.org/10.1088/0022-3727/40/2/023}

\bibitem{kozlov1995}
Kozlov K, Shepeliuk O and Samoilovich V 1995 Spatio-temporal evolution of the
  dielectric barrier discharge channels in air at atmospheric pressure {\em
  Proc. 11th Int. Conf. on Gas Discharges and their Applications (Tokyo,
  Japan)\/} vol~2 pp 142--5

\bibitem{jahanbakhsh2019correlation}
Jahanbakhsh S, Hoder T and Brandenburg R 2019 {\em Journal of Applied
  Physics\/} {\bf 126} 193305

\bibitem{janda2017}
Janda M, Hoder T, Sarani A, Brandenburg R and Machala Z 2017 {\em Plasma
  Sources Science and Technology\/} {\bf 26} 055010
  \urlprefix\url{https://dx.doi.org/10.1088/1361-6595/aa642a}

\bibitem{synek2018}
Synek P, Zem{\'a}nek M, Kudrle V and Hoder T 2018 {\em Plasma Sources Science
  and Technology\/} {\bf 27} 045008

\bibitem{becker2005}
Becker W 2005 {\em Advanced Time-Correlated Single Photon Counting
  Techniques\/} (Berlin Heidelberg: Springer)

\bibitem{itikawa2015}
Itikawa Y 2015 {\em Journal of Physical and Chemical Reference Data\/} {\bf 44}
  013105 ISSN 0047-2689 (\textit{Preprint}
  \eprint{https://pubs.aip.org/aip/jpr/article-pdf/doi/10.1063/1.4913926/14732875/013105\_1\_online.pdf})
  \urlprefix\url{https://doi.org/10.1063/1.4913926}

\bibitem{zawadzki2020}
Zawadzki M, Khakoo M~A, Voorneman L, Ratkovich L, Mašín Z, Houfek K, Dora A,
  Laher R and Tennyson J 2020 {\em Journal of Physics B: Atomic, Molecular and
  Optical Physics\/} {\bf 53} 165201
  \urlprefix\url{https://dx.doi.org/10.1088/1361-6455/ab95ef}

\bibitem{krupenie1966}
Krupenie P~H 1966 {\em The Band Spectrum of Carbon Monoxide\/} (Washington, DC:
  National Bureau of Standards NSRDS-NBS 5: National Standard Reference Series)

\bibitem{johnson1984}
Johnson M~A, Zare R~N, Rostas J and Leach S 1984 {\em The Journal of Chemical
  Physics\/} {\bf 80} 2407--2428 ISSN 0021-9606 (\textit{Preprint}
  \eprint{https://pubs.aip.org/aip/jcp/article-pdf/80/6/2407/18947733/2407\_1\_online.pdf})
  \urlprefix\url{https://doi.org/10.1063/1.446991}

\bibitem{rond2008}
Rond C, Bultel A, Boubert P and Chéron B 2008 {\em Chemical Physics\/} {\bf
  354} 16--26 ISSN 0301-0104
  \urlprefix\url{https://www.sciencedirect.com/science/article/pii/S0301010408004394}

\bibitem{mcconkey2008}
McConkey J, Malone C, Johnson P, Winstead C, McKoy V and Kanik I 2008 {\em
  Physics Reports\/} {\bf 466} 1--103 ISSN 0370-1573
  \urlprefix\url{https://www.sciencedirect.com/science/article/pii/S0370157308001488}

\bibitem{ajello1971}
Ajello J~M 1971 {\em The Journal of Chemical Physics\/} {\bf 55} 3169--3177
  ISSN 0021-9606 (\textit{Preprint}
  \eprint{https://pubs.aip.org/aip/jcp/article-pdf/55/7/3169/18876230/3169\_1\_online.pdf})
  \urlprefix\url{https://doi.org/10.1063/1.1676564}

\bibitem{itikawa2002}
Itikawa Y 2002 {\em Journal of Physical and Chemical Reference Data\/} {\bf 31}
  749--767 ISSN 0047-2689 (\textit{Preprint}
  \eprint{https://pubs.aip.org/aip/jpr/article-pdf/31/3/749/11709137/749\_1\_online.pdf})
  \urlprefix\url{https://doi.org/10.1063/1.1481879}

\bibitem{ceppelli2021}
Ceppelli M, Salden T~P~W, Martini L~M, Dilecce G and Tosi P 2021 {\em Plasma
  Sources Science and Technology\/} {\bf 30} 115010
  \urlprefix\url{https://dx.doi.org/10.1088/1361-6595/ac2411}

\bibitem{navascues2022}
Navascués P, Cotrino J, González-Elipe A~R and Gómez-Ramírez A 2022 {\em
  Chemical Engineering Journal\/} {\bf 430} 133066 ISSN 1385-8947
  \urlprefix\url{https://www.sciencedirect.com/science/article/pii/S1385894721046428}

\bibitem{bajon2024}
Bajon C, Barrate E, Sadi D, Guaitella O, Belinger A, Dap S, Hoder T and Naudé
  N 2024 {\em Plasma Sources Science and Technology\/} {\bf 33} under review

\bibitem{mrkvickova2023}
Mrkvičková M, Kuthanová L, Bílek P, Obrusník A, Navrátil Z, Dvořák P,
  Adamovich I, Šimek M and Hoder T 2023 {\em Plasma Sources Science and
  Technology\/} {\bf 32} 065009
  \urlprefix\url{https://dx.doi.org/10.1088/1361-6595/acd6de}

\bibitem{liu2019}
Liu C, Fridman A and Dobrynin D 2019 {\em Journal of Physics D: Applied
  Physics\/} {\bf 52} 105205
  \urlprefix\url{https://dx.doi.org/10.1088/1361-6463/aaf8f0}

\bibitem{kosarev2012}
Kosarev I~N, Khorunzhenko V~I, Mintoussov E~I, Sagulenko P~N, Popov N~A and
  Starikovskaia S~M 2012 {\em Plasma Sources Science and Technology\/} {\bf 21}
  045012 \urlprefix\url{https://dx.doi.org/10.1088/0963-0252/21/4/045012}

\bibitem{stepanyan2014}
Stepanyan S~A, Soloviev V~R and Starikovskaia S~M 2014 {\em Journal of Physics
  D: Applied Physics\/} {\bf 47} 485201
  \urlprefix\url{https://dx.doi.org/10.1088/0022-3727/47/48/485201}

\bibitem{bilek2023}
Bílek P, Dias T~C, Prukner V, Hoffer P, Guerra V and Šimek M 2023 {\em Plasma
  Sources Science and Technology\/} {\bf 32} 105002
  \urlprefix\url{https://dx.doi.org/10.1088/1361-6595/acf9c8}

\bibitem{brisset2019}
Brisset A, Gazeli K, Magne L, Pasquiers S, Jeanney P, Marode E and Tardiveau P
  2019 {\em Plasma Sources Science and Technology\/} {\bf 28} 055016
  \urlprefix\url{https://dx.doi.org/10.1088/1361-6595/ab1989}

\bibitem{obrusnik2018electric}
Obrusn{\'\i}k A, B{\'\i}lek P, Hoder T, \v{S}imek M and Bonaventura Z 2018 {\em
  Plasma Sources Science and Technology\/} {\bf 27} 085013

\bibitem{pancheshnyi2006}
Pancheshnyi S 2006 {\em Journal of Physics D: Applied Physics\/} {\bf 39} 1708
  \urlprefix\url{https://dx.doi.org/10.1088/0022-3727/39/8/N01}

\bibitem{hoder2016}
Hoder T, Loffhagen D, Voráč J, Becker M~M and Brandenburg R 2016 {\em Plasma
  Sources Science and Technology\/} {\bf 25} 025017
  \urlprefix\url{https://dx.doi.org/10.1088/0963-0252/25/2/025017}

\bibitem{paris2005intensity}
Paris P, Aints M, Valk F, Plank T, Haljaste A, Kozlov K and Wagner H 2005 {\em
  Journal of Physics D: Applied Physics\/} {\bf 38} 3894

\bibitem{paris2006}
Paris P, Aints M, Valk F, Plank T, Haljaste A, Kozlov K~V and Wagner H~E 2006
  {\em Journal of Physics D: Applied Physics\/} {\bf 39} 2636
  \urlprefix\url{https://dx.doi.org/10.1088/0022-3727/39/12/N01}

\bibitem{malagon2019}
Malagón-Romero A, Pérez-Invernón F~J, Luque A and Gordillo-Vázquez F~J 2019
  {\em Journal of Geophysical Research: Atmospheres\/} {\bf 124} 12356--12370
  (\textit{Preprint}
  \eprint{https://agupubs.onlinelibrary.wiley.com/doi/pdf/10.1029/2019JD030945})
  \urlprefix\url{https://agupubs.onlinelibrary.wiley.com/doi/abs/10.1029/2019JD030945}

\bibitem{pitchford2017}
Pitchford L~C, Alves L~L, Bartschat K, Biagi S~F, Bordage M~C, Bray I, Brion
  C~E, Brunger M~J, Campbell L, Chachereau A, Chaudhury B, Christophorou L~G,
  Carbone E, Dyatko N~A, Franck C~M, Fursa D~V, Gangwar R~K, Guerra V,
  Haefliger P, Hagelaar G~J~M, Hoesl A, Itikawa Y, Kochetov I~V, McEachran R~P,
  Morgan W~L, Napartovich A~P, Puech V, Rabie M, Sharma L, Srivastava R,
  Stauffer A~D, Tennyson J, de~Urquijo J, van Dijk J, Viehland L~A, Zammit M~C,
  Zatsarinny O and Pancheshnyi S 2017 {\em Plasma Processes and Polymers\/}
  {\bf 14} 1600098 (\textit{Preprint}
  \eprint{https://onlinelibrary.wiley.com/doi/pdf/10.1002/ppap.201600098})
  \urlprefix\url{https://onlinelibrary.wiley.com/doi/abs/10.1002/ppap.201600098}

\bibitem{kusyn2024}
Kusýn L, Jovanović A~P, Loffhagen D, Becker M~M and Hoder T 2024 {\em Plasma
  Sources Science and Technology\/} {\bf 33} under review

\bibitem{hagelaar2005}
Hagelaar G~J~M and Pitchford L~C 2005 {\em Plasma Sources Science and
  Technology\/} {\bf 14} 722
  \urlprefix\url{https://dx.doi.org/10.1088/0963-0252/14/4/011}

\bibitem{biagi}
Biagi database \url{www.lxcat.net} retrieved on September 12, 2023

\bibitem{sawada1972}
Sawada T, Sellin D~L and Green A~E~S 1972 {\em Journal of Geophysical Research
  (1896-1977)\/} {\bf 77} 4819--4828 (\textit{Preprint}
  \eprint{https://agupubs.onlinelibrary.wiley.com/doi/pdf/10.1029/JA077i025p04819})
  \urlprefix\url{https://agupubs.onlinelibrary.wiley.com/doi/abs/10.1029/JA077i025p04819}

\bibitem{land1978}
Land J~E 1978 {\em Journal of Applied Physics\/} {\bf 49} 5716--5721 ISSN
  0021-8979 (\textit{Preprint}
  \eprint{https://pubs.aip.org/aip/jap/article-pdf/49/12/5716/18382101/5716\_1\_online.pdf})
  \urlprefix\url{https://doi.org/10.1063/1.324589}

\bibitem{zobel1996}
Zobel J, Mayer U, Jung K and Ehrhardt H 1996 {\em Journal of Physics B: Atomic,
  Molecular and Optical Physics\/} {\bf 29} 813
  \urlprefix\url{https://dx.doi.org/10.1088/0953-4075/29/4/021}

\bibitem{kanik1993}
Kanik I, Trajmar S and Nickel J~C 1993 {\em Journal of Geophysical Research:
  Planets\/} {\bf 98} 7447--7460 (\textit{Preprint}
  \eprint{https://agupubs.onlinelibrary.wiley.com/doi/pdf/10.1029/92JE02811})
  \urlprefix\url{https://agupubs.onlinelibrary.wiley.com/doi/abs/10.1029/92JE02811}

\bibitem{bilek2018electric}
B{\'\i}lek P, Obrusn{\'\i}k A, Hoder T, \v{S}imek M and Bonaventura Z 2018 {\em
  Plasma Sources Science and Technology\/} {\bf 27} 085012

\bibitem{nishimura1968}
Nishimura H 1968 {\em Journal of the Physical Society of Japan\/} {\bf 24}
  130--143 (\textit{Preprint} \eprint{https://doi.org/10.1143/JPSJ.24.130})
  \urlprefix\url{https://doi.org/10.1143/JPSJ.24.130}

\bibitem{farley1996}
Farley D~R and Cattolica R~J 1996 {\em Journal of Quantitative Spectroscopy and
  Radiative Transfer\/} {\bf 56} 83--96 ISSN 0022-4073
  \urlprefix\url{https://www.sciencedirect.com/science/article/pii/0022407396000179}

\bibitem{tokue1990}
Tokue I, Shimada H, Masuda A, Ito Y and Kume H 1990 {\em The Journal of
  Chemical Physics\/} {\bf 93} 4812--4817 ISSN 0021-9606 (\textit{Preprint}
  \eprint{https://pubs.aip.org/aip/jcp/article-pdf/93/7/4812/18988597/4812\_1\_online.pdf})
  \urlprefix\url{https://doi.org/10.1063/1.458672}

\bibitem{hesser1968}
Hesser J~E 1968 {\em The Journal of Chemical Physics\/} {\bf 48} 2518--2535
  ISSN 0021-9606 (\textit{Preprint}
  \eprint{https://pubs.aip.org/aip/jcp/article-pdf/48/6/2518/18855935/2518\_1\_online.pdf})
  \urlprefix\url{https://doi.org/10.1063/1.1669477}

\bibitem{schlag1977}
Schlag E, Frey R, Gotchev B, Peatman W and Pollak H 1977 {\em Chemical Physics
  Letters\/} {\bf 51} 406--408 ISSN 0009-2614
  \urlprefix\url{https://www.sciencedirect.com/science/article/pii/0009261477853876}

\bibitem{maier1980}
Maier J~P and Thommen F 1980 {\em Chemical Physics\/} {\bf 51} 319--327 ISSN
  0301-0104
  \urlprefix\url{https://www.sciencedirect.com/science/article/pii/0301010480801066}

\bibitem{alderson1973}
Alderson R~J, Brocklehurst B and Downing F~A 1973 {\em The Journal of Chemical
  Physics\/} {\bf 58} 4041--4043 ISSN 0021-9606 (\textit{Preprint}
  \eprint{https://pubs.aip.org/aip/jcp/article-pdf/58/9/4041/18885489/4041\_1\_online.pdf})
  \urlprefix\url{https://doi.org/10.1063/1.1679772}

\bibitem{mackay1972}
Mackay G~I, Anglesey J~P and March R~E 1972 {\em Canadian Journal of
  Chemistry\/} {\bf 50} 2516--2517 (\textit{Preprint}
  \eprint{https://doi.org/10.1139/v72-405})
  \urlprefix\url{https://doi.org/10.1139/v72-405}

\bibitem{farley1997}
Farley D~R and Cattolica R~J 1997 {\em Chemical Physics Letters\/} {\bf 274}
  445--450 ISSN 0009-2614
  \urlprefix\url{https://www.sciencedirect.com/science/article/pii/S0009261497007124}

\bibitem{freeman1974}
Freeman C~G and Phillips L~F 1974 {\em Canadian Journal of Chemistry\/} {\bf
  52} 426--428 (\textit{Preprint} \eprint{https://doi.org/10.1139/v74-067})
  \urlprefix\url{https://doi.org/10.1139/v74-067}

\bibitem{smith1975}
Smith A~J, Read F~H and Imhof R~E 1975 {\em Journal of Physics B: Atomic and
  Molecular Physics\/} {\bf 8} 2869
  \urlprefix\url{https://dx.doi.org/10.1088/0022-3700/8/17/019}

\bibitem{smith1973}
Smith A~J, Imhof R~E and Read F~H 1973 {\em Journal of Physics B: Atomic and
  Molecular Physics\/} {\bf 6} 1333
  \urlprefix\url{https://dx.doi.org/10.1088/0022-3700/6/7/027}

\bibitem{rogers1970}
Rogers J and Anderson R 1970 {\em Journal of Quantitative Spectroscopy and
  Radiative Transfer\/} {\bf 10} 515--517 ISSN 0022-4073
  \urlprefix\url{https://www.sciencedirect.com/science/article/pii/0022407370901135}

\bibitem{sprang1977}
{van Sprang} H, Möhlmann G and {de Heer} F 1977 {\em Chemical Physics\/} {\bf
  24} 429--433 ISSN 0301-0104
  \urlprefix\url{https://www.sciencedirect.com/science/article/pii/0301010477851033}

\bibitem{qin2017}
Qin Z, Zhao J and Liu L 2017 {\em Journal of Quantitative Spectroscopy and
  Radiative Transfer\/} {\bf 202} 286--301 ISSN 0022-4073
  \urlprefix\url{https://www.sciencedirect.com/science/article/pii/S002240731730328X}

\bibitem{silva2006}
{Lino da Silva} M and Dudeck M 2006 {\em Journal of Quantitative Spectroscopy
  and Radiative Transfer\/} {\bf 102} 348--386 ISSN 0022-4073
  \urlprefix\url{https://www.sciencedirect.com/science/article/pii/S0022407306000240}

\bibitem{twist1979}
Twist J~R, Paske W~C, Rhymes T~O, Haddad G~N and Golden D~E 1979 {\em The
  Journal of Chemical Physics\/} {\bf 71} 2345--2351 ISSN 0021-9606
  (\textit{Preprint}
  \eprint{https://pubs.aip.org/aip/jcp/article-pdf/71/6/2345/18919589/2345\_1\_online.pdf})
  \urlprefix\url{https://doi.org/10.1063/1.438638}

\bibitem{naude2005electrical}
Naud{\'e} N, Cambronne J, Gherardi N and Massines F 2005 {\em Journal of
  physics D: applied physics\/} {\bf 38} 530

\bibitem{goldberg2018}
Goldberg B~M, Chng T~L, Dogariu A and Miles R~B 2018 {\em Applied Physics
  Letters\/} {\bf 112} 064102 (\textit{Preprint}
  \eprint{https://doi.org/10.1063/1.5019173})
  \urlprefix\url{https://doi.org/10.1063/1.5019173}

\end{thebibliography}

\bibliographystyle{iopart-num}

\end{document}